\documentclass[aps,twocolumn,superscriptaddress]{revtex4-1}

\usepackage{graphicx}
\usepackage{soul}
\usepackage{color}
\usepackage{bm}        
\usepackage{amsmath}
\usepackage{amssymb}   
\usepackage{array}
\usepackage[caption=false]{subfig}
\usepackage[bookmarks=false,linkcolor=blue,urlcolor=blue,colorlinks,citecolor=blue]{hyperref}
\makeatletter

\begin{document}

\title{Ising nematic quantum critical point in a metal: a 
Monte Carlo study}

\author{Yoni Schattner}
\thanks{These authors have contributed equally to this work.}
\affiliation{Department of Condensed Matter Physics, The Weizmann Institute of Science, Rehovot, 76100, Israel}

\author{Samuel Lederer}
\thanks{These authors have contributed equally to this work.}
\affiliation{Department of Physics, Stanford University, Stanford, CA 94305, USA}

\author{Steven A. Kivelson}
\affiliation{Department of Physics, Stanford University, Stanford, CA 94305, USA}

\author{Erez Berg}
\affiliation{Department of Condensed Matter Physics, The Weizmann Institute of Science, Rehovot, 76100, Israel}

\date{\today}

\begin{abstract}
The Ising nematic quantum critical point (QCP) associated with the zero temperature transition from a symmetric to a nematic {\it metal} is an
 exemplar of metallic quantum criticality.  We have carried out a minus sign-free quantum Monte Carlo study of this QCP for a two dimensional lattice model with sizes up to $24\times 24$ sites. 
For the parameters in this study, 
{ some (but not all) correlation functions} exhibit
 scaling behavior over the accessible ranges of temperature, (imaginary) time, and distance,
 and the system remains non-superconducting down to the lowest accessible temperatures. 
 The observed scaling behavior has remarkable similarities with recently measured properties of the Fe-based superconductors proximate to their putative nematic QCP.
\end{abstract}
\maketitle

\section{Introduction}

A hallmark of strongly correlated electron systems is the 
competition of ground states with different kinds of order.\cite{fradkinandme-ineluctablecomplexity}
In this context,
a central set of unsettled theoretical issues concerns the character of quantum critical points (QCPs) in 
metals~\cite{Hertz,Millis1993,conundrum}. Such metallic QCPs
have been 
 identified in several heavy fermion compounds\cite{Gegenwart2008};
evidence for quantum critical behavior has also been found in the ruthenate Sr$_3$Ru$_2$O$_7$~\cite{rost2011thermodynamics,millis-2002}, and the 
cuprate and iron-based superconductors~\footnote{Selected recent developments regarding
  quantum critical behavior in these material families include
  Refs.~\cite{Ramshaw2015} and~\cite{Fujita2014} in cuprates, and
  Refs.~\cite{hashimoto2012sharp} and~\cite{KouFisher2015} in Fe-based
  superconductors.}. 
Quantum criticality is also often invoked as a possible explanation of
the ``strange'' or ``bad'' metal
behavior seen in a variety of such materials~\cite{PALee1989,Castellani1996,Varma1997,Rosch1999,DellAnna2007,Maslov2011,Barzykin2006,Hartnoll2014}.

To date, there exists no  satisfactory theory of metallic QCPs in $d=2$ or $3$ spatial dimensions  although a number of 
field theoretic approaches have been attempted~\cite{altshuler1994low,nayak1994renormalization,nayak1994non,Altshuler1995,Abanov1999,Abanov2000,Abanov2003,Abanov2004, Lawler2006, lohneysen-2007,SSLee2009,Metlitski2010,Metlitski2010a,mross2010controlled,Maslov2010,Abrahams2011,fitzpatrick1,Dalidovich2013,Holder2015}.
Among the 
  unsettled issues are:  
  {\bf a)} the values of critical exponents 
  and which properties can be expressed as scaling functions involving these exponents, {\bf b)} the extent to which metallic QCPs are
  generically preempted
 by superconducting~\cite{Millis1992,Abanov2001Incoherent,roussev2001quantum,metlitski2010instabilities,Wang2013,Maier2014,lederer2015enhancement,metlitski2015cooper,Einenkel2015} or  other forms~\cite{metlitski2010instabilities,Efetov2012,Wang2014} of auxiliary order 
 that gap out the Fermi surface,
{\bf c)  }
whether there exists a 
``non-Fermi liquid'' metal in the quantum critical regime.
Controlled theoretical methods, that can be used both
to benchmark  the field
theories and for comparison with experiments
are greatly needed.
Determinantal quantum Monte Carlo (DQMC)\cite{Blankenbecler1981,AssaadBook}--in particular, in cases where the fermion sign problem can be circumvented~\cite{Wu2005,bergmetlitski,LiYao2015}--may serve
this purpose.  Such methods have been successfully
applied in several problems
in which
critical bosonic fluctuations are coupled to fermions with a
Dirac-like dispersion~\cite{Assaad2013,LiYao2014,WangTroyer2014,Xu2016}.

In this paper, we report a DQMC study of a two-dimensional sign problem-free lattice model that exhibits an
``Ising nematic'' QCP in a metal at finite fermion density;  in the nematic phase, the discrete lattice rotational symmetry
 is spontaneously broken from $C_4$ to $C_2$.  This is a particularly relevant QCP given that nematic order~\cite{borzi-2007,hinkov-2007,taillefer-nematic-2009, lawler2010intra,song2011direct,Riggs2015} and nematic quantum critical fluctuations~\cite{Chu2012, Gallais2013,  zhou2013quantum, Bohmer2014, Blumberg2014,  KouFisher2015} have been observed  in many of the materials
 mentioned above.

 \begin{figure}[b]
\includegraphics[width=0.9\columnwidth]{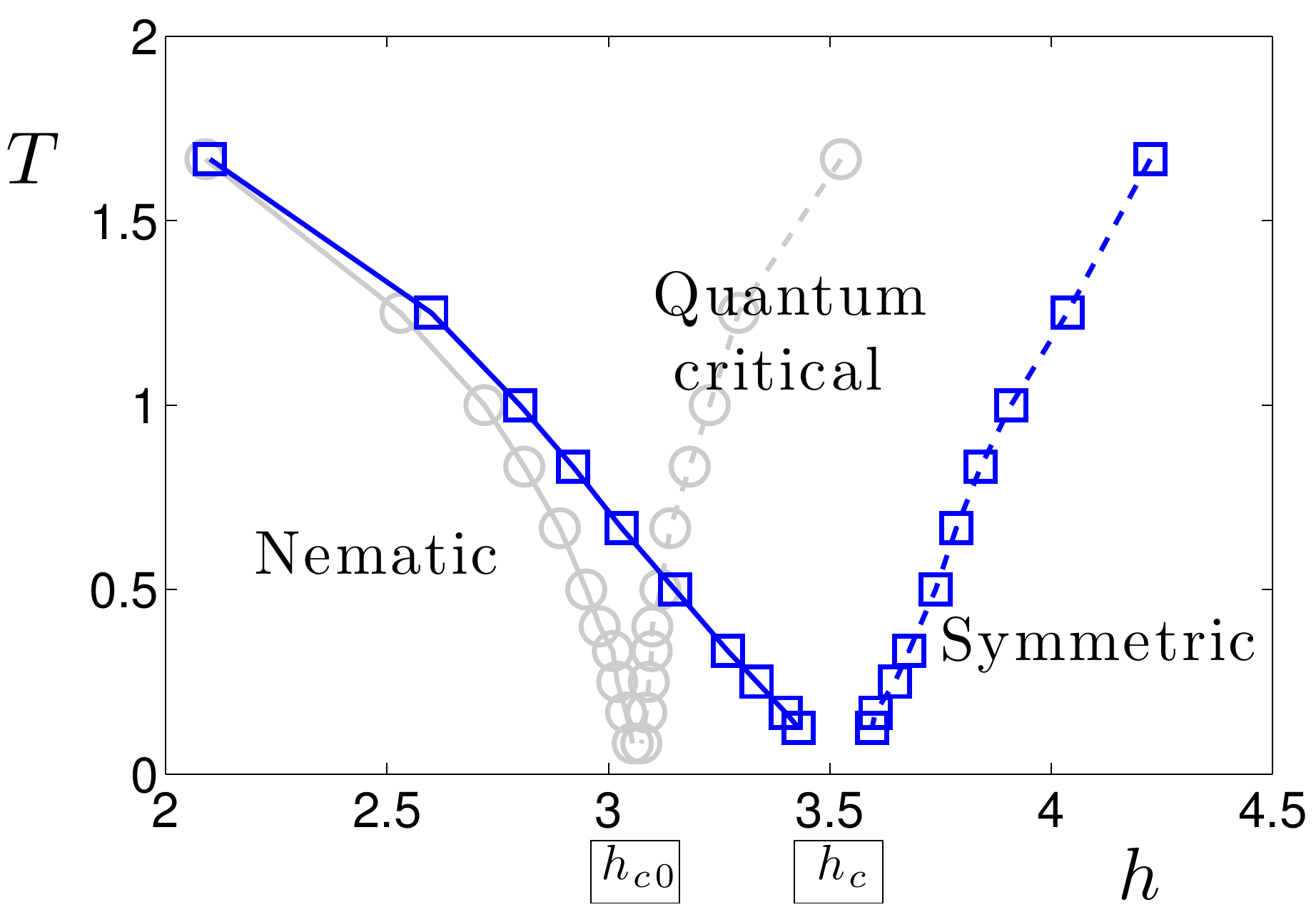}
\caption{Phase diagram of the model obtained by DQMC, as a function of the transverse field $h$ and temperature $T$. Here,
the other parameters entering the Hamiltonian Eq. \ref{eq:HFHB} are $V=t$, $\alpha = 0.5$, $\mu = -0.5t$. The solid line marks the transition temperature, $T_N$, between the nematic and the 
symmetric phases. The line extrapolates to the $T=0$ QCP at $h_c$. The dashed line marks $T_{\mathrm{cross}}$, where the nematic susceptibility reaches 50\
pale (grey) 
 lines show the 
corresponding temperatures for the case $\alpha=0$ (where the fermions and pseudospins are decoupled). In this case, the QCP occurs at $h_{c0}\approx3.06$.}
\label{fig:phase_diagram}
\end{figure}

Our simulations are limited to finite system sizes -- up to $24\times 24$ lattice sites. Within our numerical accuracy, we find evidence for a continuous nematic quantum phase transition with  critical behavior that differs significantly from that of a nematic QCP in an insulator.  Specifically,
in the disordered phase near the QCP, where the dimensionless quantum control parameter $h\geq h_c$,
the thermodynamic (zero frequency) nematic correlation function (defined in Eq. \ref{eq:chi1})
is consistent with the following 
functional description:
\begin{equation}
D(h,T,\mathbf{q},i\omega_n=0) 
\approx
\left[\frac {A} { T^{\lambda}+ b(h-h_c)+\kappa |\mathbf{q}|^{2} }\right]^\gamma
\label{eq:scaling}
\end{equation}
where $T$ is the temperature,
$\mathbf{q}$ is the wave-vector, 
$\omega_n$ is the Matsubara frequency,
and $A$, $b$, and $\kappa$ are positive dimensionful constants.
We find that the exponents in this expression take values
 $\lambda=1.0\pm 0.1$ and $\gamma=1.0\pm 0.1$. 
 (See Fig. \ref{fig:static}.) These values are in agreement with the predictions of Refs.~\cite{Millis1993,Jakubczyk2008,Bauer2011,Hartnoll2014} for the behavior of the thermodynamic susceptibility near criticality.

This implies 
that the uniform {nematic} susceptibility $\chi(h,T)\equiv 
D(h,T, \mathbf{0},0)$ 
has a Curie-Weiss form with an effective Weiss temperature
which varies linearly with $(h_c-h)$, as can be seen in Figs. \ref{fig:phase_diagram} \& \ref{fig:chi QCP}.
Eq. \ref{eq:scaling} 
{is consistent with the} scaling relation, $D(h,T,\mathbf{q},0) = \xi^{\gamma/\nu} \Phi \left(x,y\right)$,
where $\xi\sim |h-h_c|^{-\nu}$, $x=T\xi^{\tilde z}$, $y=|\mathbf{q}|\xi$.
Here, $\gamma$ is the conventionally defined susceptibility exponent, the correlation length exponent
$\nu\equiv \gamma/
 (2-\eta)$, where $\eta$ is the anomalous dimension of the nematic field, and 
we have introduced an ``apparent dynamical exponent,'' $\tilde z=(\nu \lambda)^{-1}$.  The values of these exponents derived from our Monte Carlo data are given in Table \ref{tab:exponents}. {We also include for comparison the exponents of the standard $2$-dimensional Ising QCP, which apply when the coupling to fermions is set to zero.}

Despite the accuracy of the scaling
analysis, it is important to test whether the observed behavior is characteristic of
an asymptotic quantum critical scaling regime.   To this end, we define a ``quadrupolar'' correlator, $Q$, in terms of fermion bilinears, which has the same symmetry as the nematic correlator, $D$, and therefore should have the same asymptotic critical behavior. As can be seen in Figs.~\ref{fig:comparechibeta}, \ref{fig:comparechih}, \ref{fig:comparechiq}, \ref{fig:chi_ff}, $Q$ does not obey a scaling law of the form of Eq.~(\ref{eq:scaling}) over a comparable range of $(h, T, \mathbf{q})$ as does $D$. Sufficiently close to criticality, $Q$ may be consistent with Eq.~(\ref{eq:scaling}), but the dynamic range is relatively small and the error bars on the apparent exponents are substantial.
It could be that corrections to scaling are smaller in $D$ than in $Q$, or it could be that our data simply do not reflect asymptotic quantum critical scaling.

\begin{figure}[t]
\centering
\includegraphics[width=1.0\columnwidth]{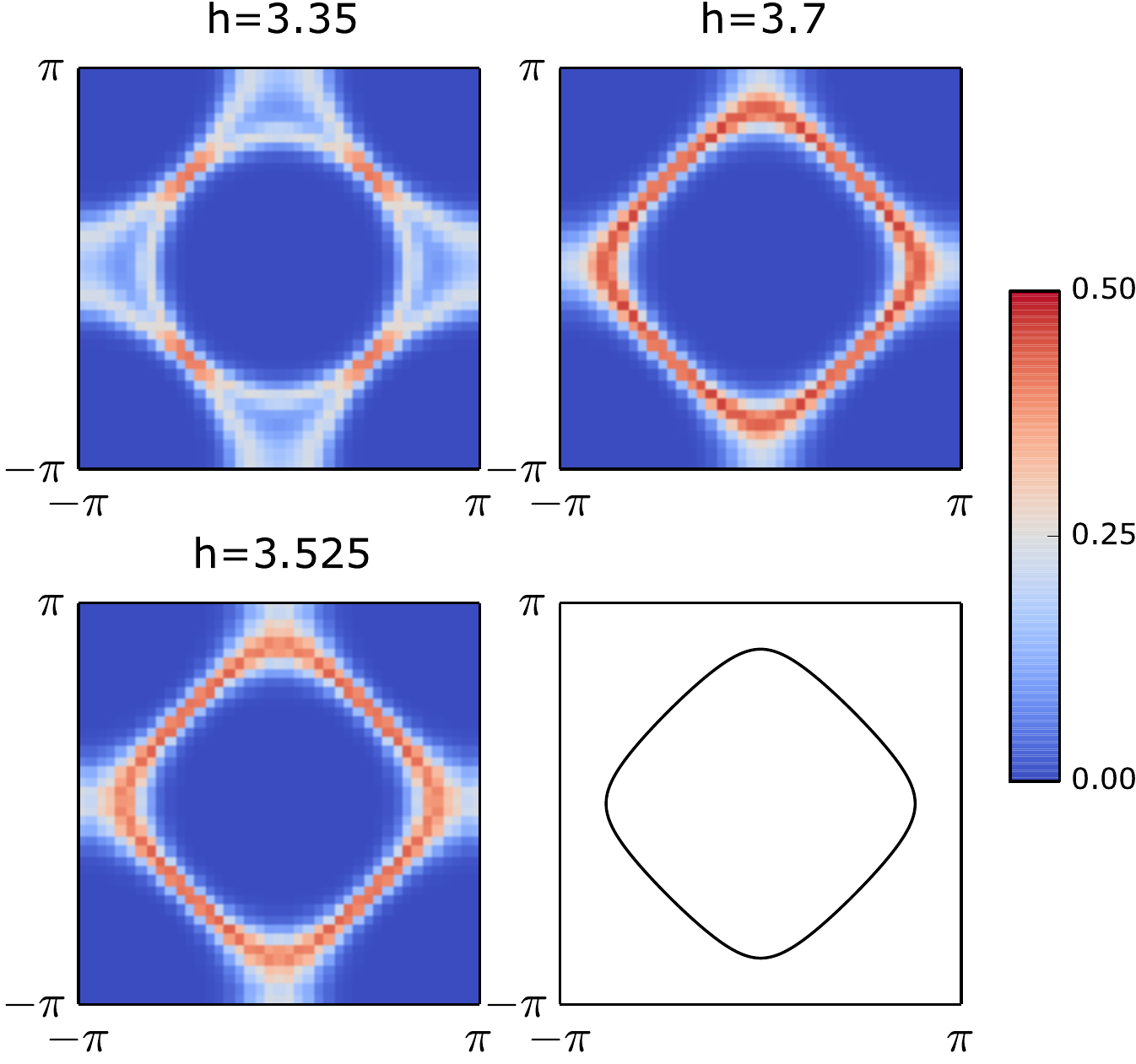}
\caption{Fermion Green's function, $G(\mathbf{k}, \tau=\beta/2)$, as a function of momentum $\mathbf{k}$ across the Brillouin zone for three different values of the transverse field: $h=3.35$ (in the nematic phase), $h=3.525$ (near the QCP), and $h=3.7$ (in the disordered phase). In the nematically ordered phase, the data is averaged over both orientations of the order parameter. $G(\mathbf{k}, \tau=\beta/2)$ is proportional to the integral
of the spectral function $A(\mathbf{k},\omega)$ over an energy window of width $T$ [see Eq.~(\ref{eq:G2})]. The temperature 
 is $T = 1/8$, and the system size is $L=20$. The data were taken from systems with either periodic or anti-periodic boundary conditions. The Fermi surfaces are seen as peaks in $G(\mathbf{k}, \tau=\beta/2)$. The lower right panel shows the Fermi surface of the bare band structure.
}
\label{fig:fermi_surface}
\end{figure}

In the fermionic 
sector we find results
consistent with  
   strongly renormalized
 Fermi liquid, 
 ``marginal Fermi liquid'' \cite{varma1989phenomenology}, or  weakly non-Fermi liquid behavior down to our lowest temperatures 
 ($T_{min} \approx 0.02 E_F$, where $E_F$ is the Fermi energy). 
In particular, at the QCP, 
 the effective quasiparticle weight, $Z_{{\bf k_F}}(T)$, defined in terms of the single-fermion Green's function in Eq.~\eqref{eq:Zdef}, 
 remains substantial.
However, it
 monotonically decreases 
 on cooling, with downward curvature.
 In a Fermi liquid, $Z_{{\bf k_F}}(T)$ would approach a positive limit as $T\to 0$, while in a weak non-Fermi liquid, it would vanish in proportion to a small power of $T$.
Fig.~\ref{fig:fermi_surface} shows maps of the low-energy spectral weight as a function of momentum in the disordered phase, at the QCP, and in the ordered phase.
 The existence\cite{Metzner2003} of ``cold-spots'' along the zone diagonals, where the quasiparticles are relatively weakly scattered, is apparent.
 Finally, no superconducting transition is found down to our lowest temperatures, although the superconducting susceptibility in the s-wave channel is peaked about $h\approx h_c$. (See Fig. \ref{fig:P_of_T}.)

\begin{table}
\protect\caption{Critical exponents for a $2d$
Ising QCP with $z=1$ (classical $3d$ Ising), and
scaling exponents  from  DQMC simulations near a metallic Ising nematic QCP.
In the limit of vanishing coupling to the fermions, exponents from  DQMC simulations are consistent with $3d$ classical Ising values.
$\tilde z$ is the ``apparent dynamical exponent'' defined 
below Eq.~(\ref{eq:scaling}). 
}
\begin{center}
 \begin{tabular}{ | m{3cm} || c | c | c | c| }
\hline
Critical Exponents               & $\nu$            & $\gamma$    & $\eta$  &$  \tilde z      $        \\  \hline
$2d$  Ising QCP              & $0.63$          & $1.24$          & $0.04$  &$   1.0  $          \\  \hline
$2d$ {\it metallic}  Ising nematic QCP  & $0.5\pm0.1$ & $1.0\pm0.1$ & $0.0\pm 0.3$ & $2.0\pm 0.3$  \\  \hline
  \end{tabular}
\end{center}
\label{tab:exponents}
\end{table}
\label{sec:model}

The remainder of the paper is organized as follows: in 
Sec. \ref{sec:model} we 
define the lattice Hamiltonian and describe its phase diagram; in the next section, we provide evidence for the statements above regarding 
thermodynamic nematic and quadrupolar correlations (\ref{sec:bose}),  dynamical nematic and quadrupolar correlations (\ref{sec:dynamics}), superconductivity (\ref{sec:SC}), and single-fermion correlations (\ref{sec:fermionic}); finally, in 
Sec. \ref{sec:discussion}, we discuss various caveats concerning the interpretation of our results,  and their 
bearing 
on both prior theoretical work and the interpretation of experiments.

\section{Model and phase diagram}

\begin{figure}[b]
\centering
\includegraphics[width=0.65\columnwidth]{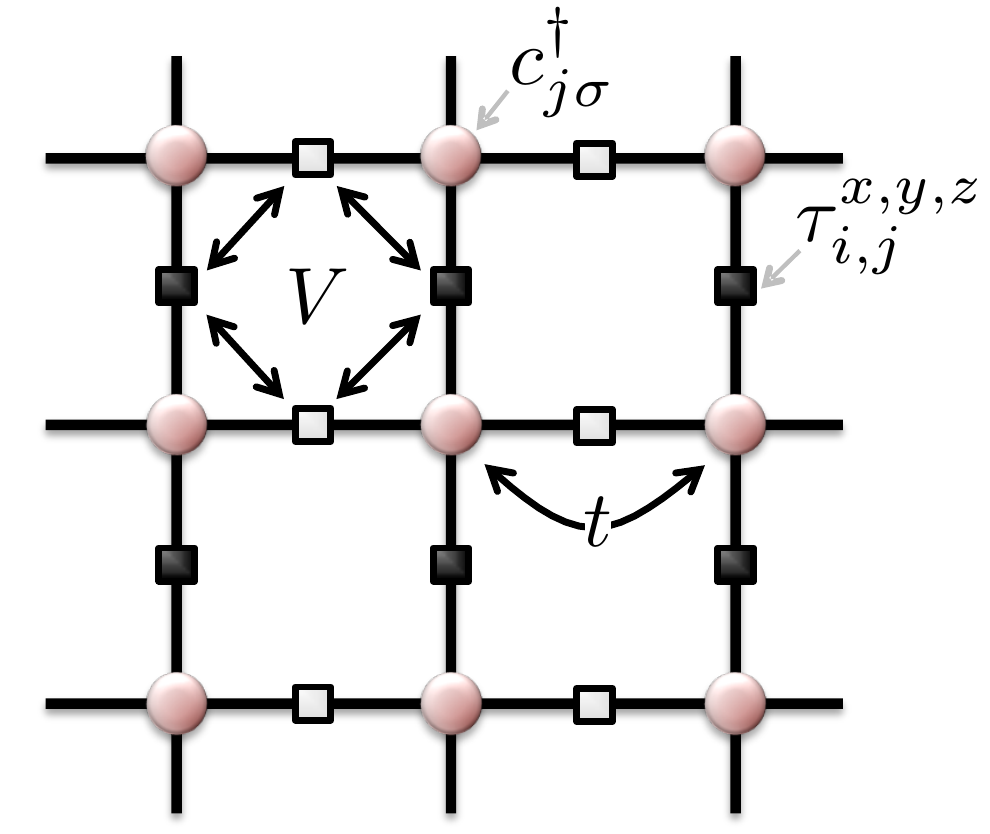}
\caption{Illustration of the lattice model (\ref{eq:H}). Spinful electrons reside on the sites, while the Ising pseudospins 
live on the bonds. The pseudospins interact with their neighbors antiferromagnetically, and are coupled to the fermion bond density.}
\label{fig:model}
\end{figure}

Our model is illustrated in Fig.~
\ref{fig:model}, and is defined on a two-dimensional square lattice. Every lattice site hosts a single (spinful) fermionic degree of freedom. Each link has
a pseudospin-$1/2$ degree of freedom that couples to the fermion bond-density. The system is described by the following Hamiltonian:
\begin{equation}
H = H_f + H_b + H_\mathrm{int},
\label{eq:H}
\end{equation}
where
\begin{eqnarray}
H_f &=& -t\sum_{\langle i,j \rangle, \sigma} c^\dagger_{i\sigma} c^{\vphantom{\dagger}}_{j\sigma} - \mu \sum_{i, \sigma} c^\dagger_{i\sigma} c^{\vphantom{\dagger}}_{i\sigma}, \nonumber \\
H_b &=& V\sum_{\langle \langle i,j\rangle;\langle k,l \rangle \rangle} \tau^z_{i,j} \tau^z_{k,l} -h \sum_{\langle i,j\rangle } \tau^x_{i,j}, \nonumber \\
H_\mathrm{int} &=& \alpha t \sum_{\langle i,j\rangle, \sigma} \tau^z_{i,j} c^\dagger_{i\sigma} c^{\vphantom{\dagger}}_{j\sigma}.
\label{eq:HFHB}
\end{eqnarray}
Here, $c^\dagger_{j\sigma}$ creates a fermion on site $j$ with spin $\sigma=\uparrow,\downarrow$, $\langle
i,
j \rangle$ denotes a pair of nearest-neighbor sites on the square lattice, $t$ and $\mu$ are the hopping strength and chemical potential, respectively, $\tau^a_{i,j}$ ($a = x,y,z$) denote Pauli matrices that act on the
pseudospin that lives on the bond connecting the neighboring sites $i$ and $j$, $V>0$ is the nearest-neighbor Ising interaction between neighboring pseudospins (here,
$\langle \langle i,j\rangle;\langle k,l \rangle \rangle$ denotes a pair of nearest-neighbor bonds), 
h is the strength of a transverse field that acts on the pseudospins, and $\alpha$ is the
dimensionless coupling strength between the pseudospin and the fermion bond density.

The pseudospins are not related to the physical spins of the electrons; 
their ordering corresponds to a nematic transition. The model~(\ref{eq:H}) should be viewed as an effective lattice model, designed to give a nematic QCP. Microscopically, the nematic degrees of freedom could
{represent (via Hubbard-Stratonovich transformation)
interactions involving the same electrons that form the Fermi surface, or
another, independent degree of freedom (such as a phonon mode
that becomes} soft at a structural transition). {
So long as
the properties of the
} QCP are universal, the low-energy behavior does not depend on its microscopic origin.

For $\alpha = 0$, the system is composed of two decoupled sets of degrees of freedom: free fermions, 
and pseudospins governed by $H_b$, which has the form of a $d=2$ transverse field Ising model. At zero temperature,
the pseudospins undergo a second-order quantum phase transition from a paramagnet to an ``antiferromagnet'' that breaks $90^\circ$ rotational symmetry at $h=h_{c0}$.  This transition is
in the 
$3$ dimensional classical Ising universality class. At $T=0$, the pseudospin degrees of freedom are gapped in both the nematic and isotropic phases. At finite temperatures, a line of second-order classical $d=2$ Ising transitions 
extends from the QCP in the $h-T$ plane.

For 
non-zero $\alpha$,
the phase diagram is similar, but exhibits quantitative and qualitative modifications.
Fig.~\ref{fig:phase_diagram}
shows the phase diagram, obtained by DQMC, for both $\alpha=0$ and $\alpha \ne 0$. The $\alpha\ne 0$ transition 
 between the nematic and isotropic phases remains second order, and extrapolates to a new QCP, shifted relative to 
 $h_{c0}$.  More striking is the change in the slope with which the 
phase boundary,
$T_N(h)$, approaches the QCP. For $\alpha=0$, the slope diverges at low temperature, consistent 
with the expectation for 
the transverse field Ising transition, 
where $T_N \propto |h-h_{c0}|^{\nu z}$ with $\nu={0.63}$ and $z=1$. In contrast, for $\alpha>0$, we find that $T_N \propto (h_c-h)$. On the disordered side of the transition,
we define a crossover line  by identifying $h_{\mathrm{cross}}(T)$ as the value of $h$ at which the 
nematic susceptibility at fixed $T$ has fallen to half of its value at $h=h_c$.
$T_{\mathrm{cross}}(h)$, the  inverse of $h_{\mathrm{cross}}$, also vanishes linearly with $h$ upon approaching the QCP, although its slope is 
steeper than that of $T_N(h)$.

\begin{figure}[t]
\centering
\includegraphics[width=0.8\columnwidth]{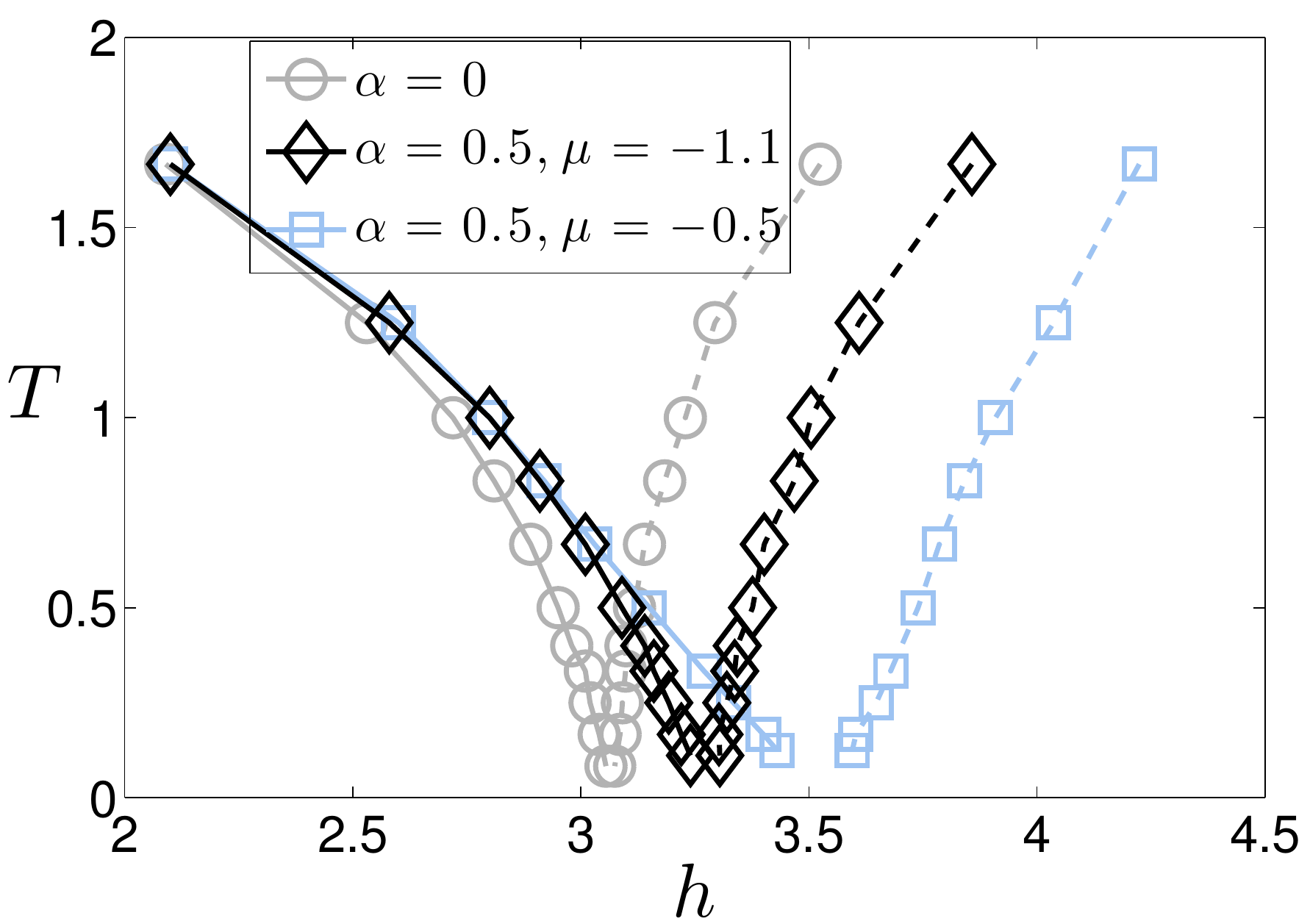}
\caption{The phase diagram for several sets of parameters. Gray circles represent the decoupled problem ($\alpha=0$), while black diamonds and blue squares represent the coupled problem  ($\alpha=0.5$) at $0.6$ and $0.8$ fermions per site, respectively.}
\label{phasediagram}
\end{figure}

The linear behavior of
{$T_N(h)$ for small $h_c-h$} is also seen for other model parameters.   In Fig.~\ref{phasediagram}
 we show the phase diagram for
two values of the fermion density, controlled by the chemical potential $\mu$. As the fermion density is
 reduced, both $h_c-h_{c0}$ and the
 range over which $T_N(h)$ is linear become smaller, indicating that the
 effect of the coupling between electrons near the Fermi surface and the nematic modes becomes weaker. The fact that
 $T_N(h)$ appears linear at low temperature for both values of fermion density is consistent with this being a \emph{universal} property of the metallic QCP.

\section{DQMC Results and analysis}
\label{sec:DQMC}

We have simulated the Hamiltonian~(\ref{eq:H}) with system sizes between $L=8$ and $L=24$, and temperatures between $0.025t$ and $5t$. These simulations do not suffer from the minus sign problem: the
fermion determinants obtained from integrating out the spin up and spin down fermions 
are identical and real,  
so the product is nonnegative.  
Global updates of the pseudospin space-time configurations 
are introduced in order to overcome critical slowing down.

Some technical details related to the DQMC simulations are discussed in Appendix~\ref{app:DQMC}. Here, we shall focus on the results and data analysis. Unless stated otherwise, the model parameters used in the simulations were: $V=t$, $\alpha = 0.5$, $\mu = -0.5t$, $t=1$.

\begin{figure}[t]
\hspace{-1.0cm}
\centering
\includegraphics[width=1.1\columnwidth]{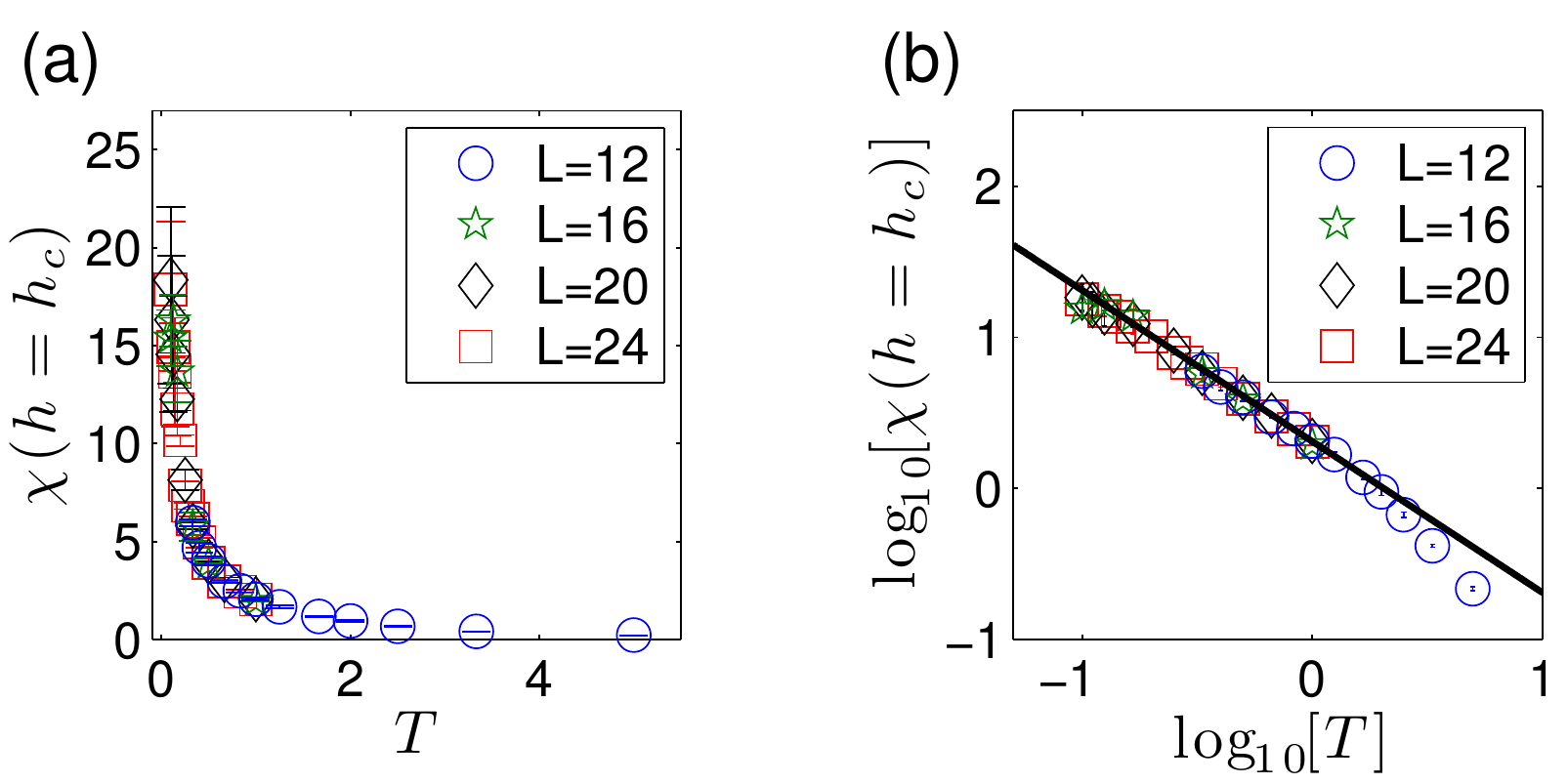}
\caption{Nematic susceptibility at 
{the critical $h$}, $\chi(
 h_c, T)$, as a function of temperature. Panel a) shows the susceptibility on a linear scale. 
 The same data are displayed on a log-log scale in panel b). The black line is proportional to $1/T$.}
\label{fig:chi QCP}
\end{figure}

\subsection{Thermodynamic
correlations}
\label{sec:bose}
The two-point function for the nematic order parameter field is defined as
\begin{equation}
D (h,T,\mathbf{q},i\omega_n)  =
\frac{1}{L^2} \sum_{i,j} \int_0^\beta d\tau e^{i\omega_n \tau -i\mathbf{q}\cdot \mathbf{r}_{ij}}
\langle N_{i}(\tau) N_{j}(0) \rangle.
\label{eq:chi1}
\end{equation}
Here, $\mathbf{r}_{ij} = \mathbf{r}_i - \mathbf{r}_j$ 
and the nematic order parameter is defined as $N_i = \sum_j \eta_{ij} \tau^z_{ij}$,
with $\eta_{ij}=1/4$ for $\mathbf{r}_{ij} = \pm \hat{\mathbf{x}}$, $\eta_{ij}=-1/4$ for $\mathbf{r}_{ij} = \pm \hat{\mathbf{y}}$, and $\eta_{ij}=0$ otherwise. The phase diagram discussed above was determined using the 
nematic susceptibility 
 and standard finite size scaling techniques, described in Appendix~\ref{app:finite}.
 We also define a quadrupolar order parameter, ${\cal N}_i = \sum_{j,\sigma} \eta_{ij} (c^\dagger_{i,\sigma}c_{j,\sigma}+c^\dagger_{i,\sigma}c_{i,\sigma})$ in terms of which the quadrupolar correlation function is
\begin{equation}
Q (h,T,\mathbf{q},i\omega_n)  =
\frac{1}{L^2} \sum_{i,j} \int_0^\beta d\tau e^{i\omega_n \tau -i\mathbf{q}\cdot \mathbf{r}_{ij}}
\langle {\cal N}_{i}(\tau) {\cal N}_{j}(0) \rangle.
\label{eq:chiQ}
\end{equation}
Note that, while microscopically quite different, $N$ and ${\cal N}$ have the same symmetries, and both develop anomalous expectation values in the nematic phase.

Fig.~\ref{fig:chi QCP} shows $\chi(h=h_c, T)$,
which diverges upon cooling.
 Examining the data on a log-log plot, [Fig.~\ref{fig:chi QCP}(b)], the susceptibility is seen to follow a power law, $\chi(h=h_c,T) = A /T^{\lambda}$ with $\lambda\approx 1$, over more than a decade of temperature between $0.1V<T\lesssim 2V$. The slope of $\log(\chi)$ vs. $\log(T)$ 
is independent of the system size for $L\geq 12$.

\begin{figure}
\includegraphics[width=1.1\columnwidth]{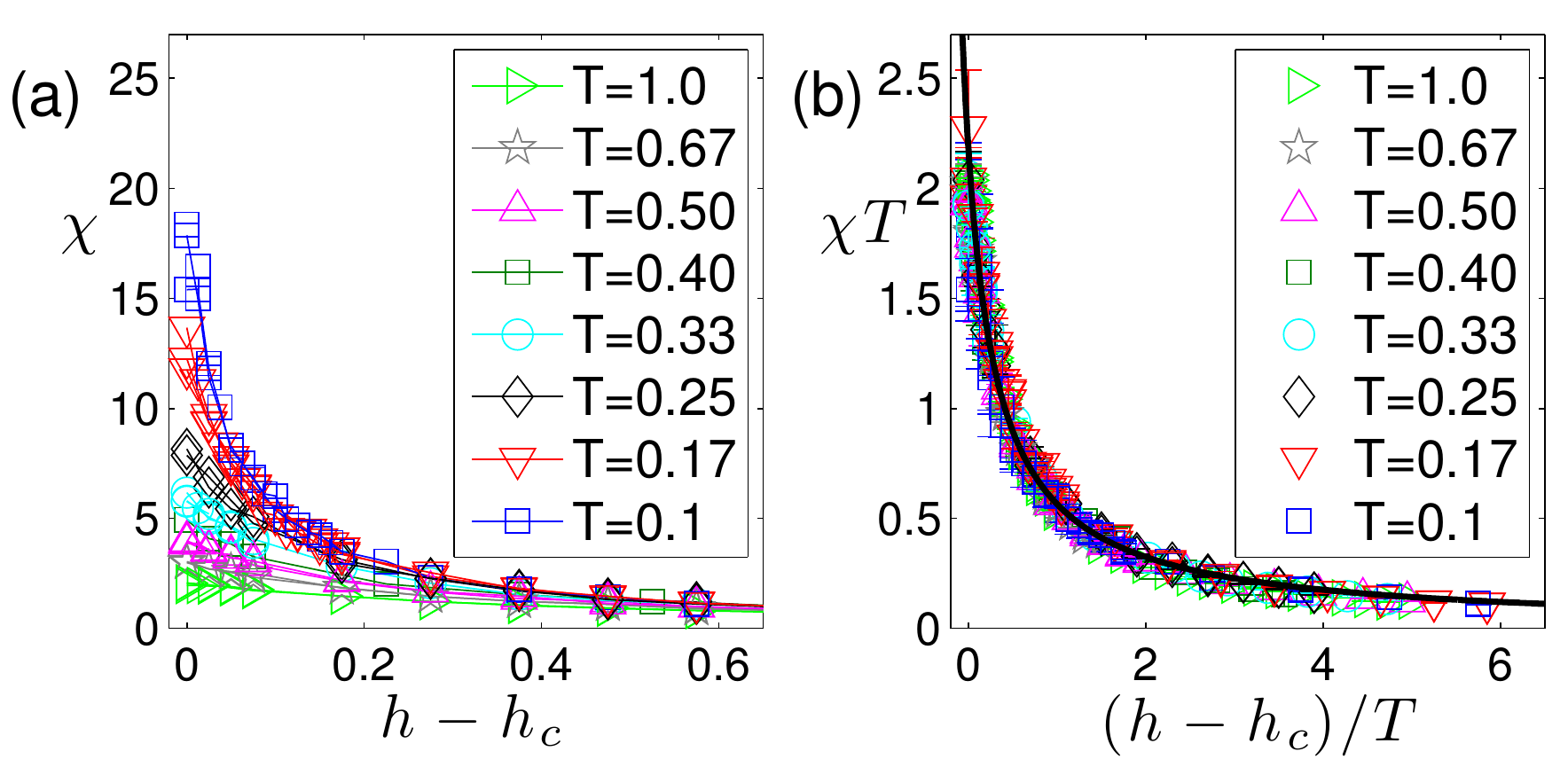}
\caption{(a) Nematic susceptibility, $\chi(h,T)$, as a function of $h-h_c$ for system sizes $L=16,20$, and $24$ at various temperatures. (b) The same data, scaled appropriately for $\lambda=1$ and $\gamma=1$, 
collapse on a single curve. 
The black curve is the function $F(x) = 2.1/(1+2.9x)$.}
\label{fig:chi h T}
\end{figure}

\begin{figure}
\includegraphics[width=1.0\columnwidth]{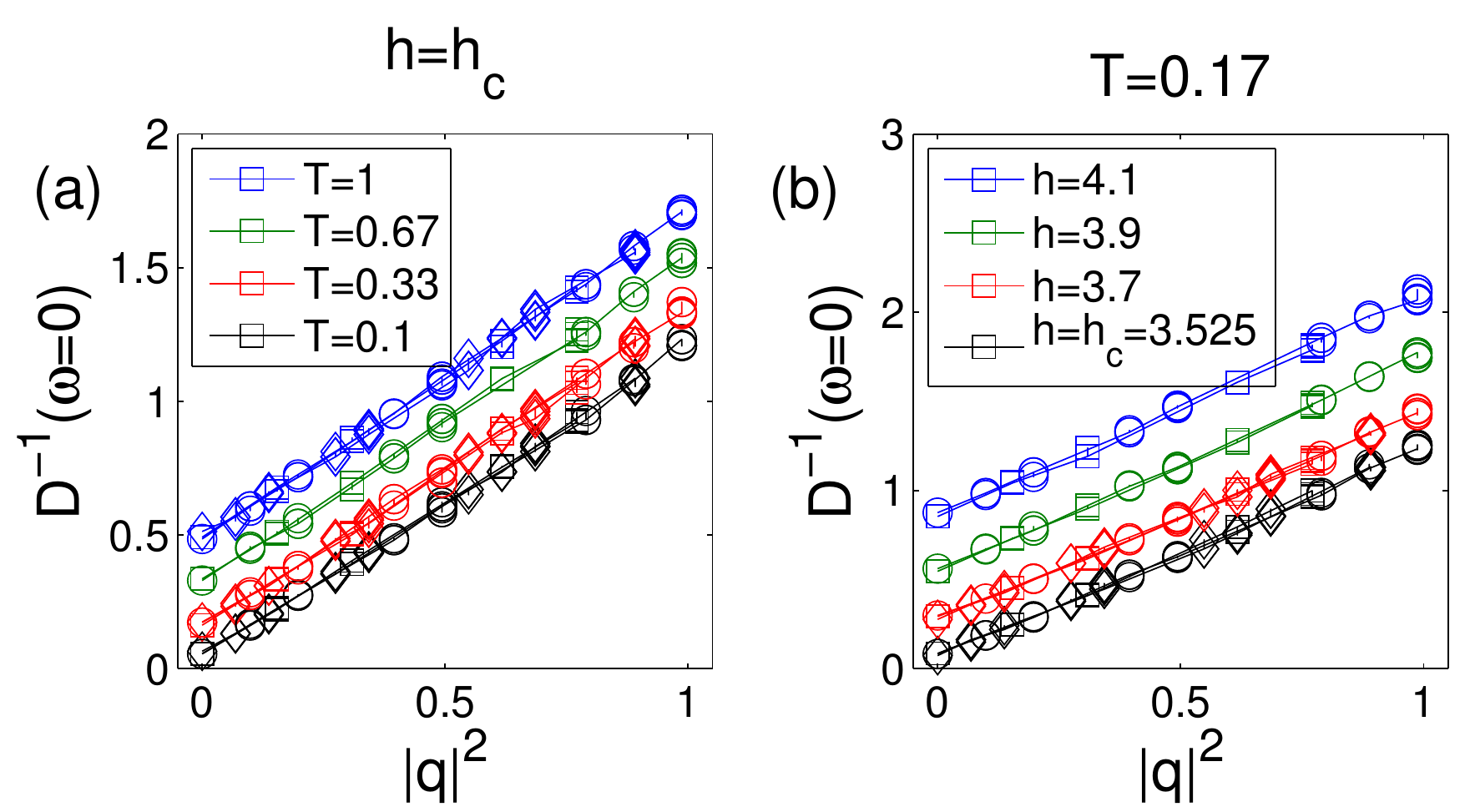}
\caption{Quadratic momentum dependence of the inverse {nematic correlator} at zero frequency. (a) shows $D^{-1}$ vs $|{\bf q}|^2$ for all orientations of $\mathbf{q}$, at $h=h_c$ and various temperatures. 
(b) shows the same for fixed temperature $T=0.17 t$ and a variety of values of the quantum tuning parameter $h$.  
In each case the momentum dependence remains quadratic. 
Momenta from multiple system sizes (represented by squares for $L=16$, circles for $L=20$, and diamonds for $L=24$) fall on the same curve, indicating that the properties shown are in the thermodynamic limit.}
\label{fig:chi h T k}
\end{figure}

\begin{figure}
\includegraphics[width=0.9 \columnwidth ]{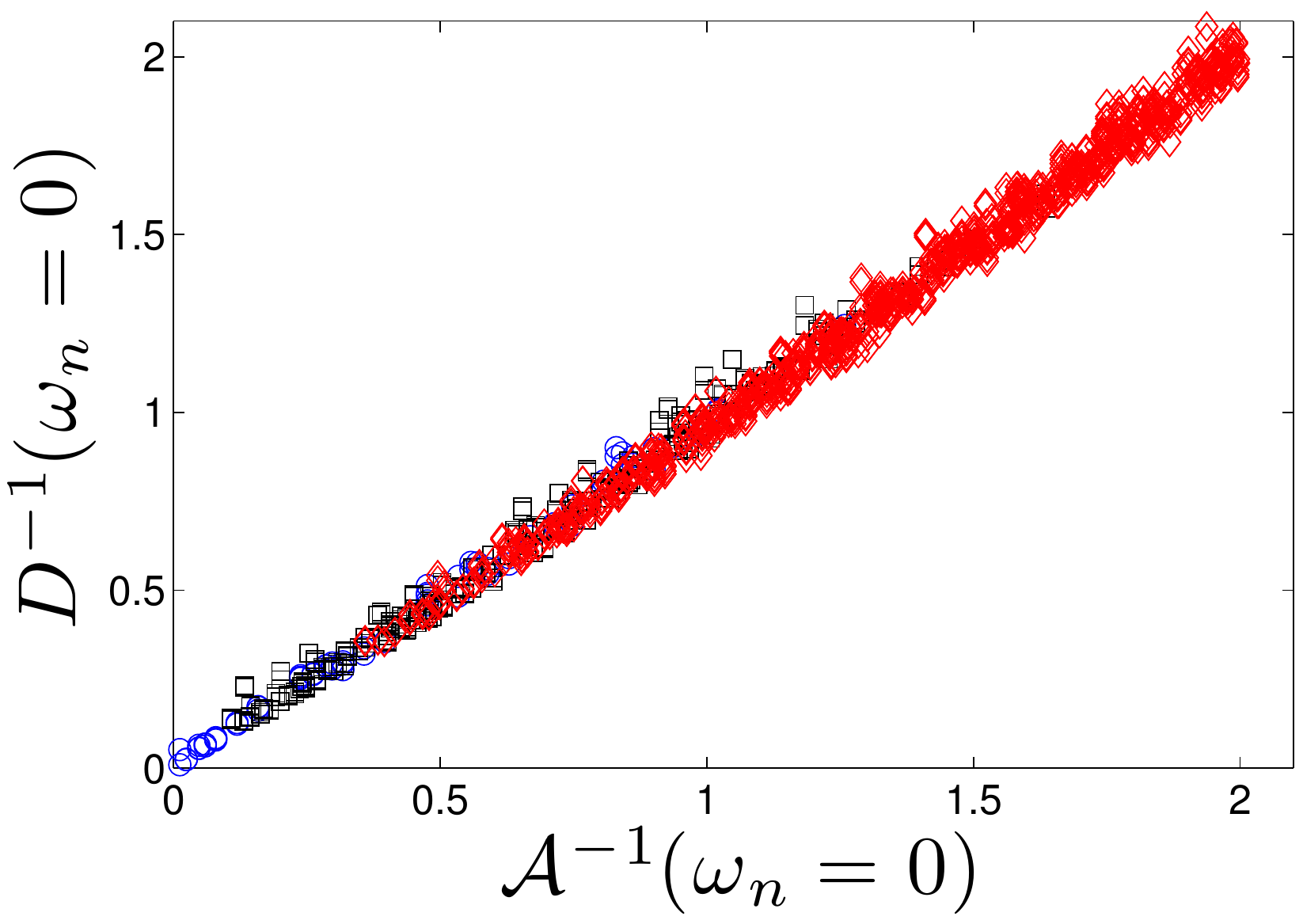}
\caption{
Comparison between the thermodynamic ($\omega_n=0$) functional approximant, ${\cal A}$ from Eq.~\ref{eq:D_dynamics}, and the nematic correlator, $D$.
Data shown represent $L=16$, 20,  and 24, 
$h=3.525$, 3.7, 3.9, and 4.1, and 
 temperatures $T=1.0$, 0.67, 0.5, 0.33, 0.25, 0.17, 0.13, 0.1,  0.05, and 0.025. 
To exhibit the momentum dependence we have used  
blue circles for  $\mathbf{q}=\mathbf{0}$ (71 data points), black squares for $|\mathbf{q}|=(2\pi/L)$ and $(2\pi/L)\sqrt{2}$ (568 data points), and red diamonds for all other $\mathbf{q}$ (1916 data points).
}
\label{fig:static}
\end{figure}

\begin{figure}
\includegraphics[width=1.0\columnwidth]{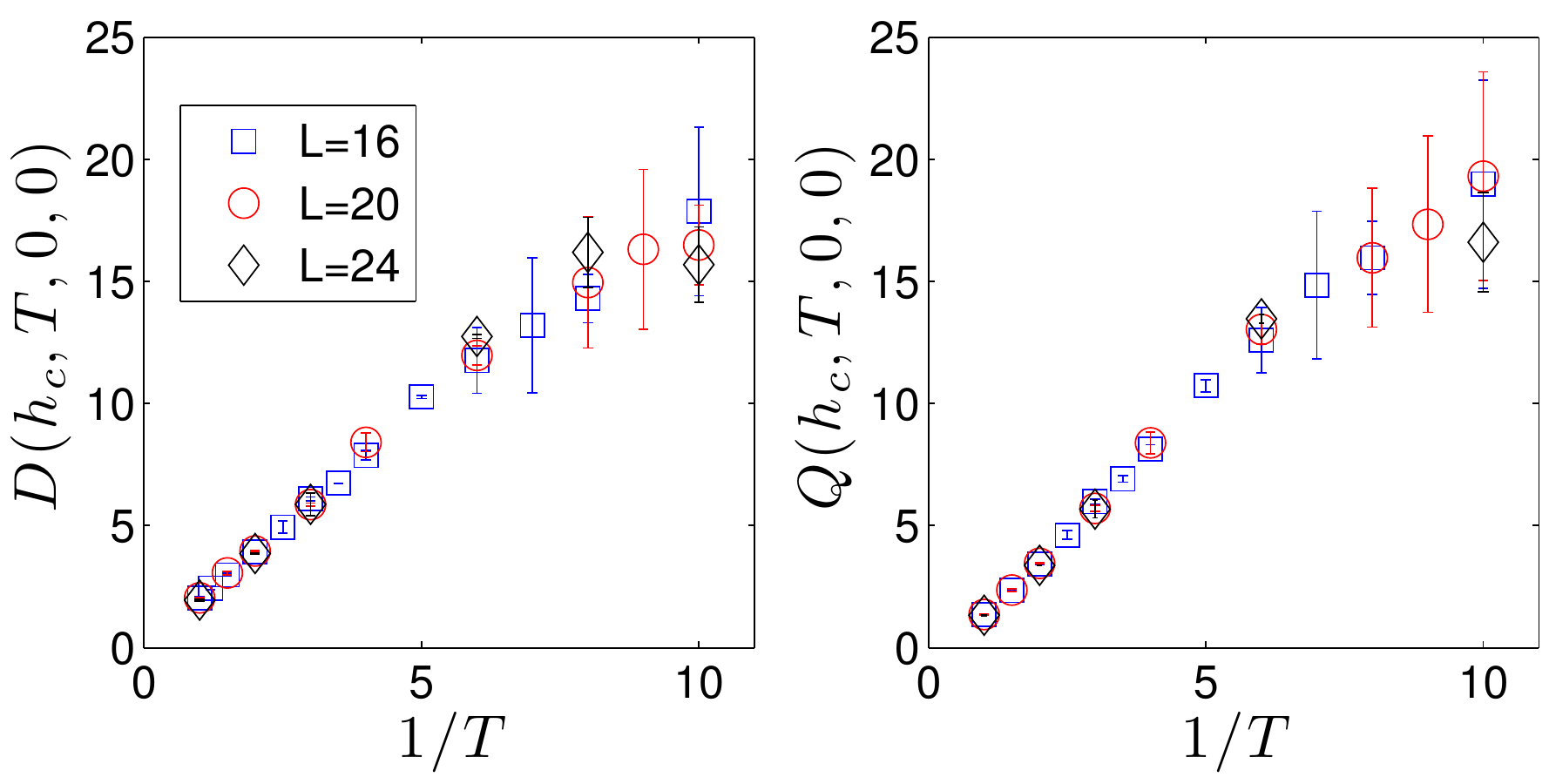}
\caption{Nematic and quadrupolar susceptibilities vs inverse temperature at $h=h_c$.  Multiple system sizes  are shown with squares for $L=16$, circles for $L=20$, and diamonds for $L=24$. The error bars reflect uncertainties due to finite system sizes, estimated from the differences between the measured susceptibilities of systems with different boundary conditions.}
\label{fig:comparechibeta}
\end{figure}
\begin{figure}
\includegraphics[width=1.0\columnwidth]{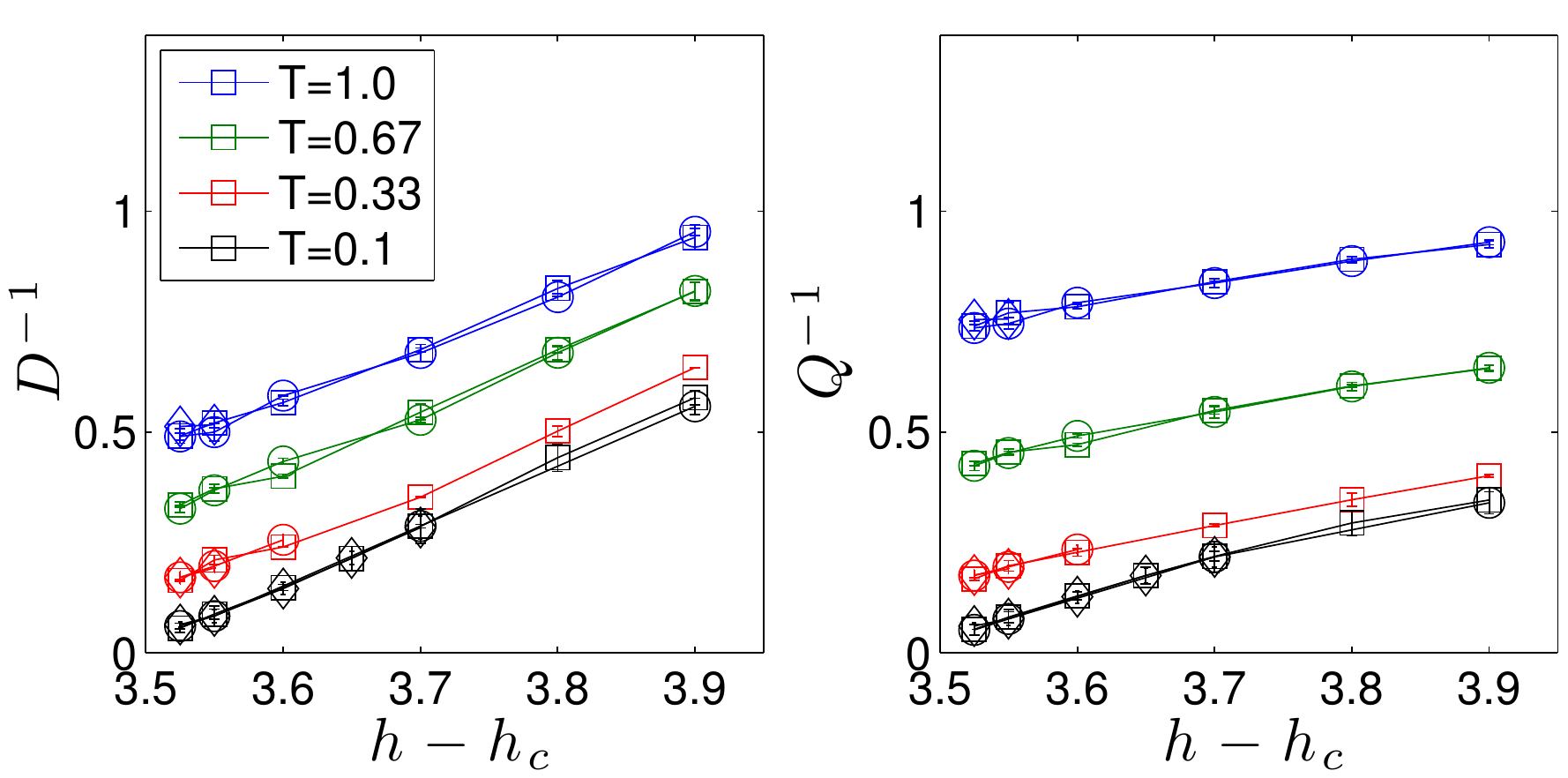}
\caption{Inverse nematic and quadrupolar susceptibilities vs $h-h_c$ at various temperatures.  Multiple system sizes  are shown with squares for $L=16$, circles for $L=20$, and diamonds for $L=24$. The error bars reflect uncertainties due to finite system sizes, estimated from the differences between the measured susceptibilities of systems with different boundary conditions.}
\label{fig:comparechih}
\end{figure}
\begin{figure}
\includegraphics[width=1.\columnwidth]{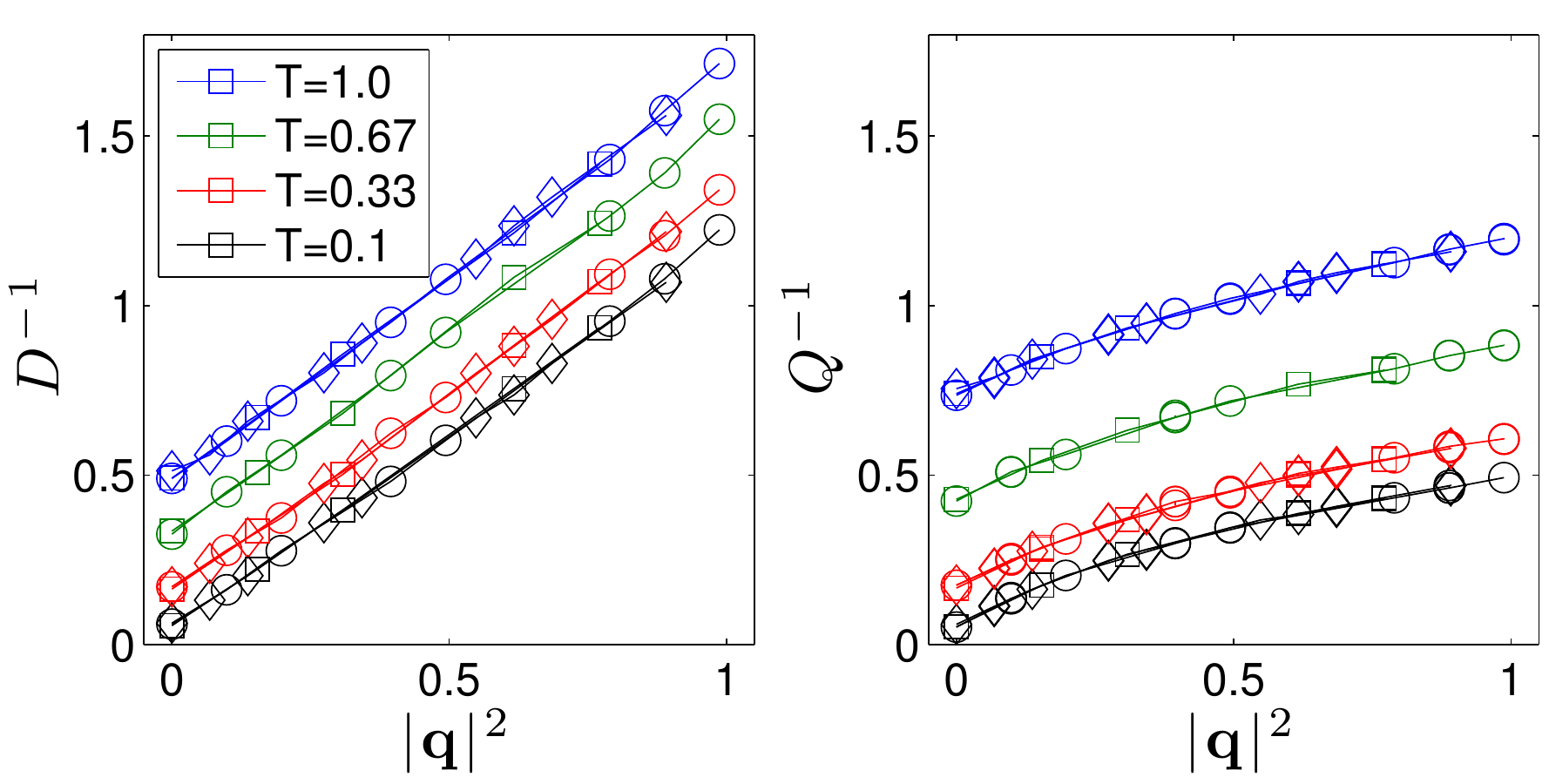}
\caption{Inverse nematic and quadrupolar correlators vs $|\mathbf{q}|^2 $
 for all orientations of $\mathbf{q}$ at $h=h_c$ and $T$ from $0.1$ to $1.0$.
Momenta from multiple system sizes (represented by squares for $L=16$, circles for $L=20$, and diamonds for $L=24$) fall on the same curve, indicating that the properties shown are in the thermodynamic limit. Error bars are smaller than the symbol size.}
\label{fig:comparechiq}
\end{figure}

To establish the scaling behavior of $\chi$ away from $h_c$,
we plot the susceptibility for different temperatures and system sizes, as a function of $h-h_c$ [Fig.~\ref{fig:chi h T}(a)]. For the system sizes displayed ($16\leq L \leq 24$), $\chi$ is only weakly dependent on $L$. In Fig.~\ref{fig:chi h T}(b) we plot $T^{\lambda\gamma}\chi(h,T)$ for $h>h_c$ as a function of $(h-h_c)/T^{\lambda}$, with $\lambda=1.0$ and $\gamma=1.0$. The different curves collapse onto each other, 
as expected for a scaling function. 

Next, we examine the momentum dependence of the static correlator $D(h,T,\mathbf{q},i\omega_n=0)$ near the QCP. Fig.~
\ref{fig:chi h T k} shows $D^{-1}(h,T, \mathbf{q}, i\omega_n = 0)$ as a function of $|\mathbf{q}|^2$  for
various values of $h-h_c$ and $T$.
For small momentum, the momentum dependence consists of an essentially isotropic term, approximately proportional to $|\mathbf{q}|^2$.
Combined with the 
$h$ and $T$ dependences discussed above, we deduce that the static nematic {correlator}
is well-described by 
 Eq. \eqref{eq:
scaling}, 
with the aforementioned exponents, and with
$A \approx 2.1 $, $b\approx 2.9 t$, $\kappa\approx 2.6 t$,
 and the lattice spacing set to unity. This is explicitly demonstrated in Fig. \ref{fig:static}.  (In the figure, ${\cal A}(h,T,\mathbf{q},\omega_n)$
 is the functional approximant to $D(h,T,\mathbf{q},\omega_n)$ , defined in Eq. \ref{eq:D_dynamics}.)

{In order to evaluate the consistency of the scaling form, Eq.~(\ref{eq:scaling}), we have also measured the quadrupolar correlator, $Q$, over a similar range of parameters as $D$. }
In Fig. \ref{fig:comparechibeta}, we compare the nematic and the quadrupolar susceptibilities at $h=h_c$ as a function of 
temperature.
Like $D$, $Q$ varies approximately linearly as a function of $1/T$, although there are large error bars at low temperatures.
In Fig. \ref{fig:comparechih} we show the $h$ dependence of $D^{-1}$ and $Q^{-1}$ at various temperatures.  In both cases, the $h$ dependence is approximately linear for small $h-h_c$. However, the range of $h-h_c$ in which $D^{-1}$ is linear is larger than the corresponding range of linearity of $Q^{-1}$, and the slope of $D^{-1}$ at small $h-h_c$ is approximately $T$ independent, while for $Q^{-1}$ it is noticeably temperature dependent.

The momentum dependence of the two correlators is shown in Fig. \ref{fig:comparechiq}.  Both appear isotropic, depending only on $|\mathbf{q}|^2$. However, $Q^{-1}$, in contrast to $D^{-1}$, has noticeable downward curvature. $D^{-1}$ depends on temperature through an essentially momentum-independent shift, while the temperature dependence of $Q^{-1}$ is more complicated. For more details see Appendix E.

\begin{figure}
\includegraphics[width=1.0\columnwidth ]{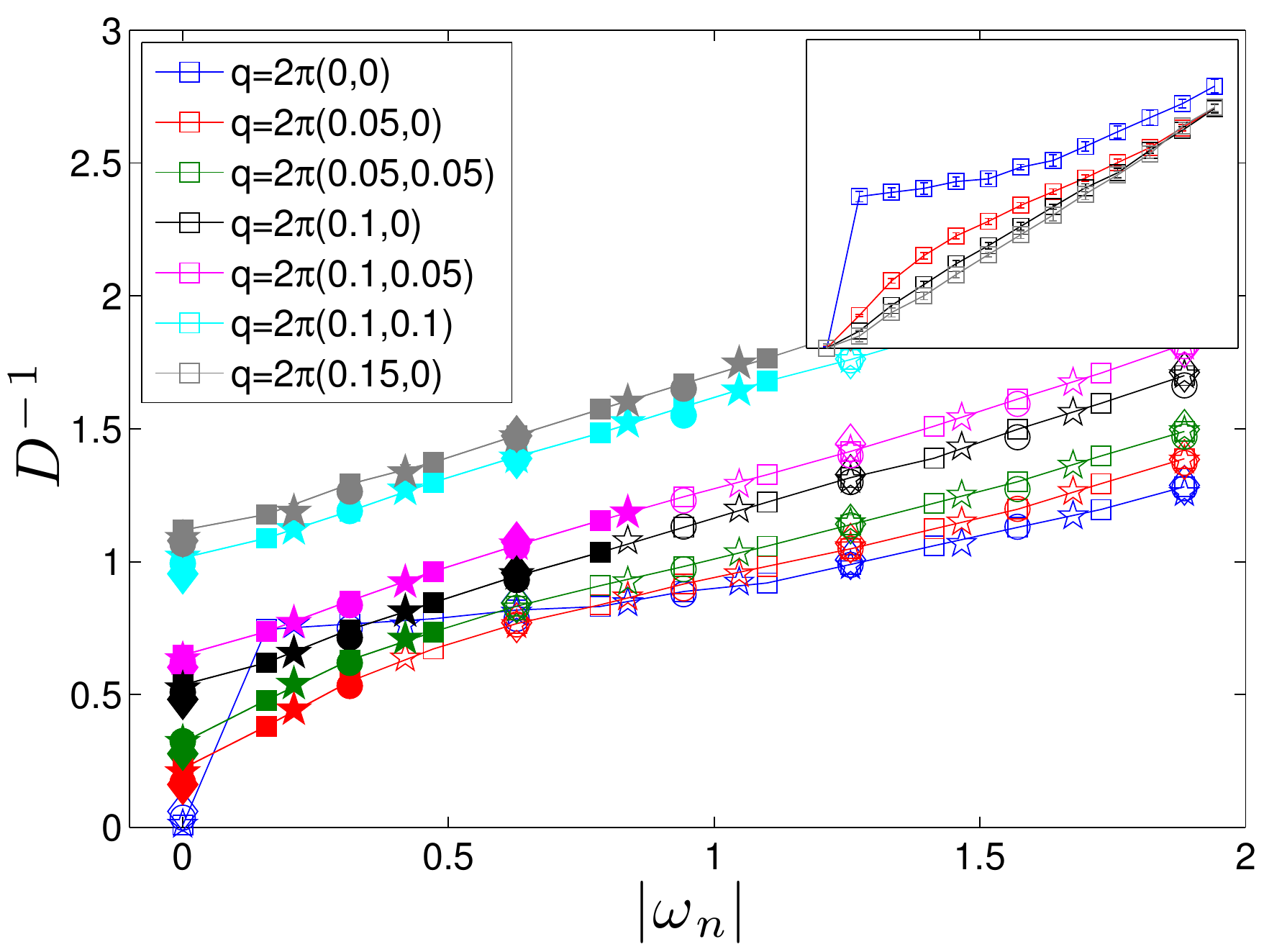}
\caption{Frequency dependence of the inverse {nematic correlator} at $h\approx h_c$ for a variety of small momenta $\mathbf{q}$ in an $L=20$ system at temperatures $T=0.025, 0.033,0.05,0.1$ (shown 
as squares, stars, circles, and diamonds respectively). Different colors represent different values of $\mathbf{q}$, and filled symbols mark frequencies
$|\omega_n| < v_F |{\bf q}|$
where $v_F$ is the minimum value
of the bare Fermi velocity.
The frequency dependence
is essentially linear except at the few smallest frequencies and momenta. The inset shows
the subset of the data with $T=0.025$ and momenta along the $(10)$ direction
shifted by their zero frequency value, with the same scale as the main figure.
 }
\label{fig:chi w k}
\end{figure}

\begin{figure}
\includegraphics[clip=true,trim= 0 0 0 0,width=\columnwidth ]{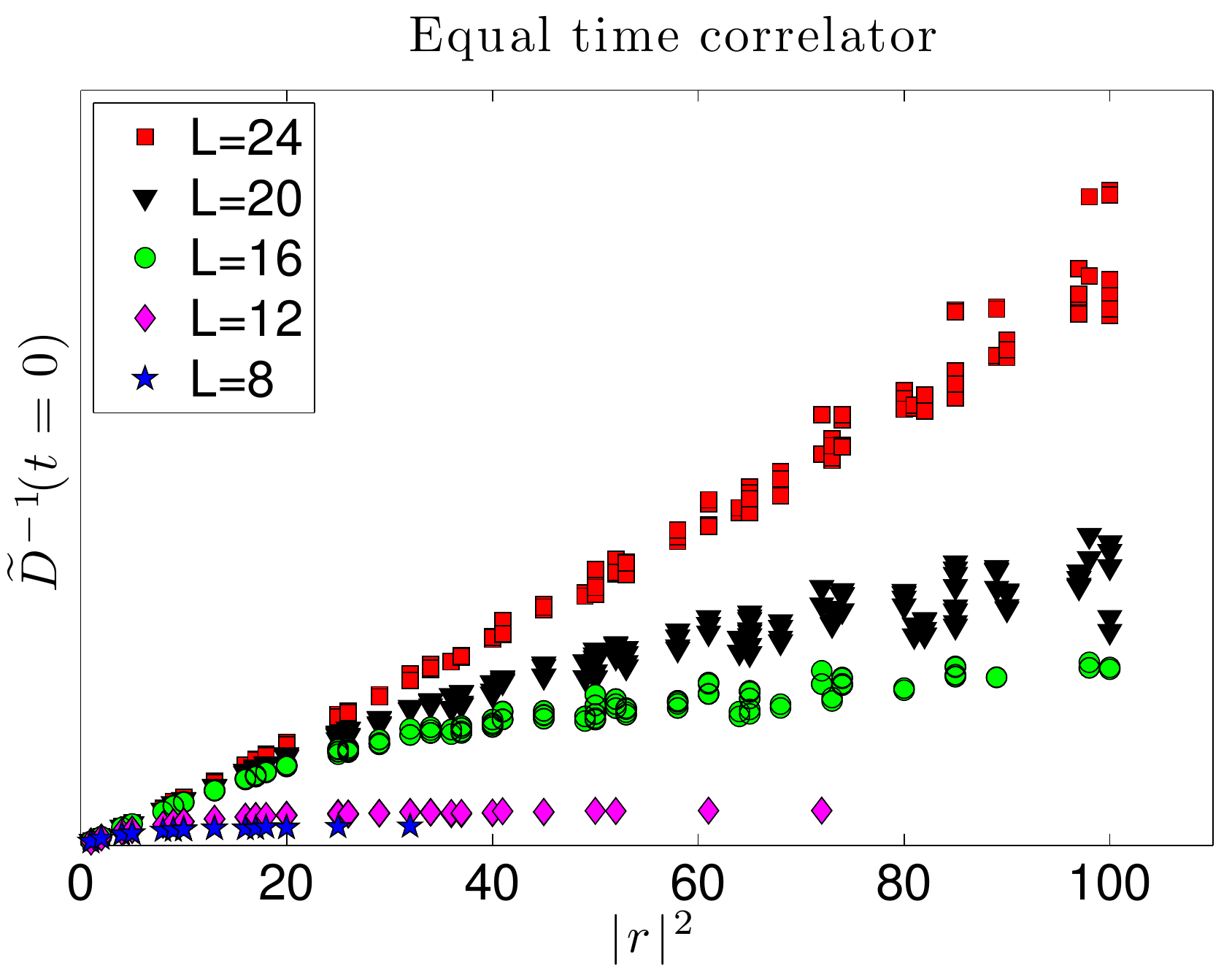}
\caption{The inverse of the equal time 
 nematic correlator at $h=h_c$ and $T=0.1$, plotted versus the square of the spatial separation ${\bf r}$ for various system sizes. The narrow spread in the data (which have not been chosen to lie along any high symmetry direction) indicates an emergent isotropy of the 
 correlations in this regime. 
 { The apparent $1/|{\bf r}|^2$ behavior 
  in the thermodynamic limit is 
  what one would get by Fourier transform of ${\cal A}$ from Eq. \ref{eq:D_dynamics}.}}
\label{fig:chi r}
\end{figure}

\subsection{Dynamic 
{correlations}}
\label{sec:dynamics}

\begin{figure}
\includegraphics[width=1.0\columnwidth ]{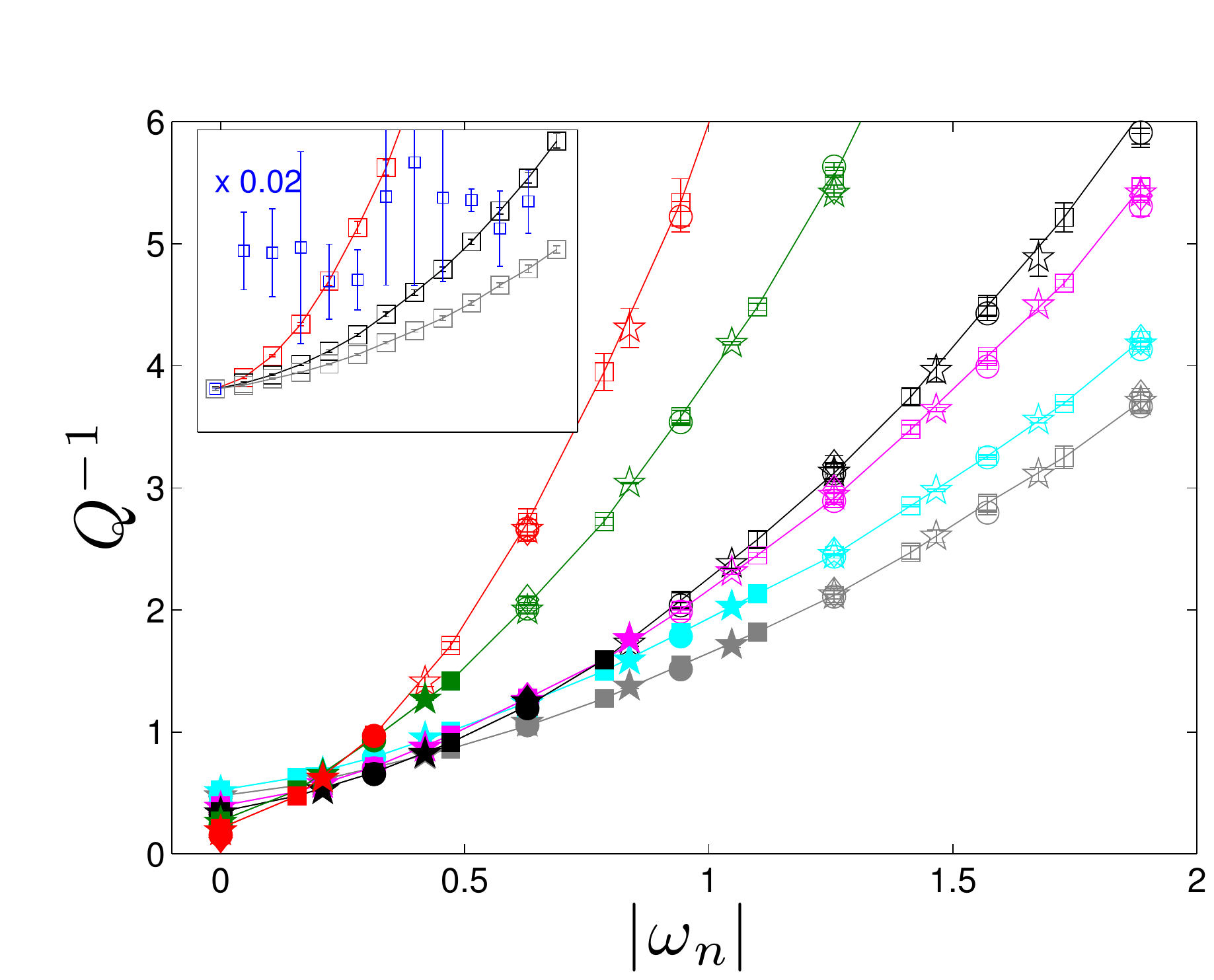}
\caption{Frequency dependence of the inverse quadrupolar  correlator at $h\approx h_c$ for a variety of small momenta 
 in an $L=20$ system at temperatures $T=0.025, 0.033,0.05,0.1$ (shown 
as squares, stars, circles, and diamonds respectively). Different colors represent different values of $\mathbf{q}$, with the same color scale as in
Fig. \ref{fig:chi w k}.  Filled symbols mark frequencies $|\omega_n| < v_F |{\bf q}|$, where $v_F$ is the minimum value
of the bare Fermi velocity.  The inset shows the subset of the data with $T=0.025$ and momenta along the $(10)$ direction shifted by their zero frequency value, with the same scale as the main figure. The data for $\mathbf{q}=0$ are scaled down by a factor of $0.02$ to appear on the same scale.}
\label{fig:chi_ff}
\end{figure}

The dependence of
$D$ on Matsubara frequency, $\omega_n=2\pi T n$, is shown in Fig. \ref{fig:chi w k}.
At intermediate $\omega_n$ ($0.5t \lesssim \omega_n \lesssim 2t$) and for $\mathbf{q}\ne \mathbf{0}$, $D^{-1}$ is an approximately linear function of $|\omega_n|$ with a slope that is independent of $T$, and also independent of both the direction and magnitude of $\mathbf{q}$.  However, at the smallest 
non-zero momenta, a different {frequency dependence is visible  for $|\omega_n|<0.5t$. This is emphasized in the inset of Fig.~\ref{fig:chi w k}, where $D^{-1}(\mathbf{q},i\omega_n=0)$ has been subtracted.} The frequency dependence at $\mathbf{q} = \mathbf{0}$ is different from the behavior seen for non-zero $\mathbf{q}$; for $\omega_n \lesssim t$, $D^{-1}(\mathbf{q}=0,i\omega_n)$
appears to saturate to a frequency independent value, followed by a sudden drop at $\omega_n=0$. The frequency dependence of $D^{-1}$ away from the critical point, shown in Fig.~\ref{fig:chiwk_gapped} in the Appendix \ref{app:dump}, is similar to the behavior of $D^{-1}$ at criticality, apart from a constant positive shift (that corresponds to the fact that we are on the disordered side of the transition, and hence $D(\mathbf{q}=0,i\omega=0)$ remains finite in the thermodynamic limit).

The bulk of the data at  small values of $h-h_c$, $T$, $\mathbf{q}$, and $\omega_n$  is well described
 by the simple  ``functional approximant'' $D(h,T, \mathbf{q}, i \omega_n) \approx {\cal A}(h,T, \mathbf{q}, i \omega_n)$, where
\begin{equation}
\label{eq:D_dynamics}
{\cal A}(h,T, \mathbf{q}, i \omega_n)\equiv
\frac A {T+b(h-h_c)+\kappa |\mathbf{q}|^2 +c|\omega_n|}.
\end{equation}
As already shown in Fig.~\ref{fig:static}, this approximant is extremely accurate for all the data at $\omega_n=0$.
As can be seen from  Fig.~\ref{fig:chi huge} in Appendix E, it does an equally good job for the dynamical response for $\mathbf{q}$ not too small, but fails to describe the  observed dynamics for the few lowest values of $|\mathbf{q}|$.

The equal time correlator gives complementary information about the dynamics, since
it integrates over all frequencies.
{Moreover,
``equal time'' is the same in both real and imaginary time.  } Thus, in Fig. \ref{fig:chi r} we exhibit the inverse of the equal time {correlator},
$\tilde D(\tau=0,\mathbf{r})$ for $h=h_c$ at $T=0.1 t$, as a function of distance for various values of $L$.  Note that for the largest size system ($L=24$), $\tilde D(0,\mathbf{r}) \sim 1/|\mathbf{r}|^2$, consistent with what one would get by Fourier transform of $
{\cal A}$.
$\tilde D$ is shown for all orientations of $\mathbf{r}$, so the narrow spread of points 
{shows} the high degree of isotropy exhibited by the data.  Note also that for relatively small $|\mathbf{r}|$, the data is independent of $L$, at least for $L\geq 16$, but that at longer distances the data for the smaller systems deviate systematically from the $|\mathbf{r}|^{-2}$ behavior.

The frequency dependence of  $Q$ is shown in Fig.~\ref{fig:chi_ff}.  It is clearly qualitatively different from the behavior of $D$ in Fig. \ref{fig:chi w k}, and correspondingly cannot be approximated by any expression of the form in Eq. \ref{eq:D_dynamics}. 
The uniform quadrupolar density is nearly conserved, as is apparent from the very large values of $Q^{-1}(\mathbf{q}=\mathbf{0})$ at nonzero frequencies, shown in the inset.
\subsection{Superconductivity}
\label{sec:SC}

\begin{figure}[t]
\centering
\includegraphics[width=1.0\columnwidth]{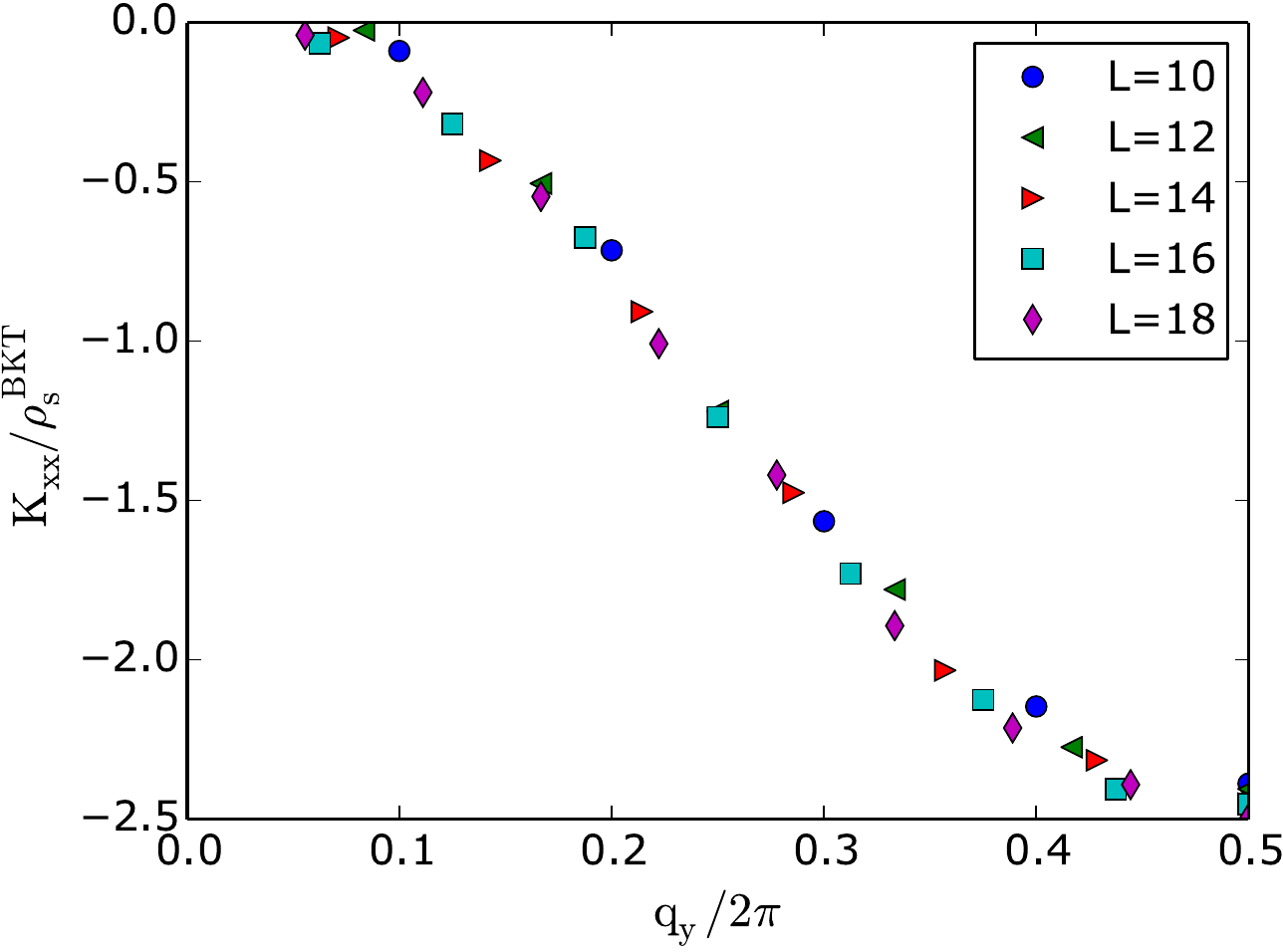}
\caption{
$K_{xx}(q_x=0,q_y)$ 
 defined in 
 Eq.~(\ref{eq:Kxx}) 
 for $T=0.05t$, $h=h_c$, and various values of system size, normalized by the universal value $\rho_s^{BKT}$. 
 Momenta from different system sizes lie on the same curve, illustrating the thermodynamic limit. At small values of $q_y$, $K_{xx}(q_x=0,q_y)$ 
  lies well below $\rho_s^{BKT}$, indicating the system is not superconducting.}
\label{fig:rho_s}
\end{figure}

\begin{figure}[t]
\centering
\includegraphics[width=1.0\columnwidth]{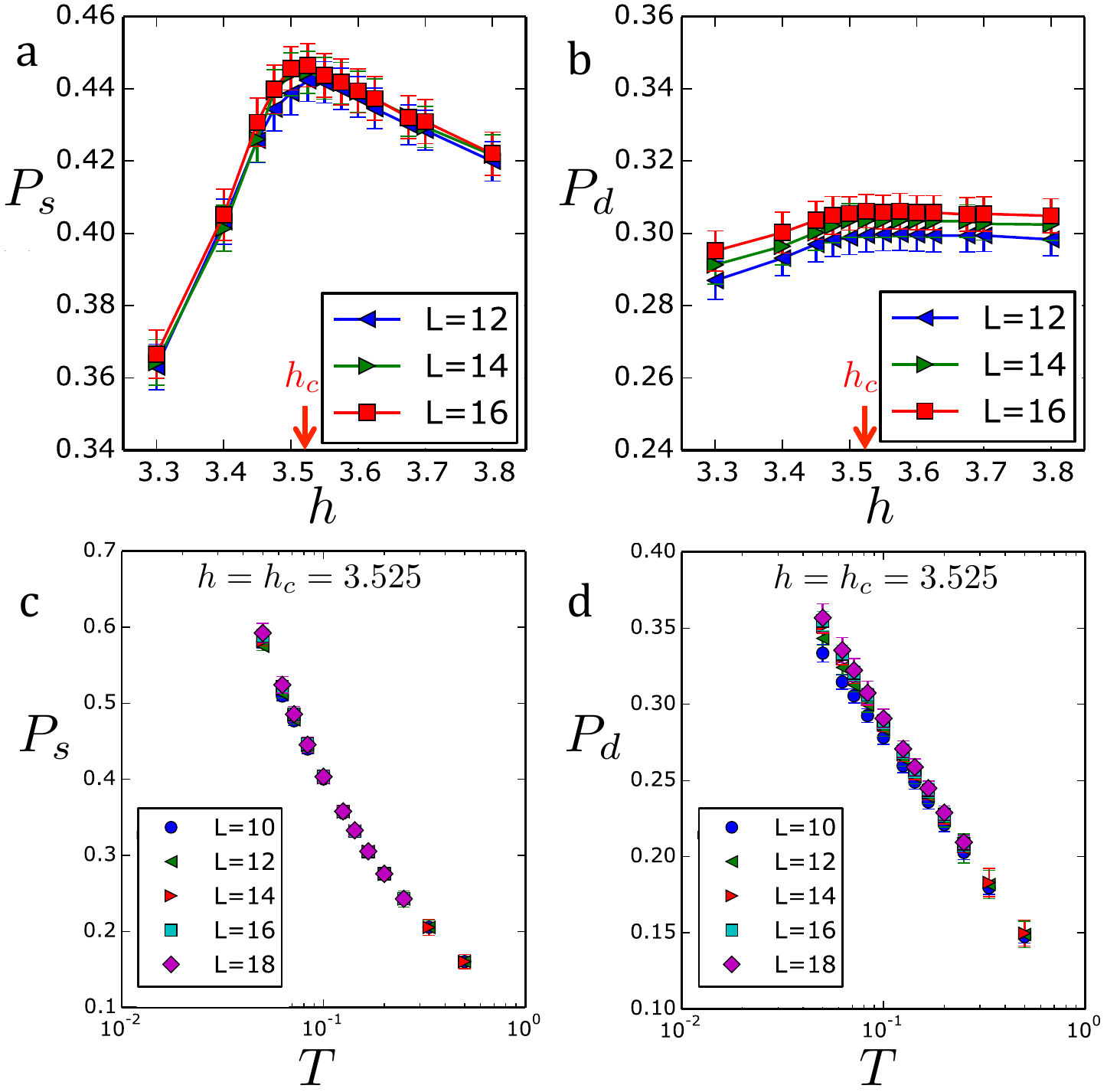}
\caption{(a,b) Pairing susceptibility in the $s$--wave and $d$--wave channels, $P_{s,d}$ [Eq.~(\ref{eq:sc_suscept})], as a function of $h$, at $T=0.083$. Different system sizes are displayed. (c,d) $P_{s,d}$ as a function of temperature at the QCP, $h = 3.525$. }
\label{fig:P_of_T}
\end{figure}

We have probed for superconducting correlations in the vicinity of the QCP by measuring the superfluid density and the superconducting susceptibility, as a function of $h$ and $T$. The superfluid density is given by~\cite{Scalapino1993, Paiva2004}
\begin{equation}
\rho_s = \lim_{q_y\rightarrow 0} \lim_{L\rightarrow \infty} K_{xx}(q_x=0, q_y) 
\label{eq:rho_s}
\end{equation}
where
\begin{equation}
K_{xx}(\mathbf{q}) \equiv \frac{1}{4} \left[\Lambda_{xx} (q_x\rightarrow 0, q_y=0)  -  \Lambda_{xx} (\mathbf{q})  \right],
\label{eq:Kxx}
\end{equation}
and $\Lambda_{xx}$ is the current-current correlator
\begin{equation}
\Lambda_{xx}(\mathbf{q}) = \sum_{i}\int_0^\beta d\tau e^{-i\mathbf{q} \cdot \mathbf{r}_i} \langle j_x(\mathbf{r}_i,\tau) j_x(0,0)\rangle.
\end{equation}
Here, the current density operator is given by $j_x(\mathbf{r}_i) = \sum_\sigma it(1 + \alpha \tau^z_{i,j}) c^\dagger_{i\sigma} c^{\vphantom{\dagger}}_{j\sigma} + \mathrm{H.c.}$,  where $\mathbf{r}_j = \mathbf{r}_i + \hat{x}$. $K_{xx}(\mathbf{q})$ describes the response of the system to a static orbital magnetic field $B(\mathbf{q})$:
\begin{equation}
j_x(\mathbf{q}) = - 4 K_{xx}(\mathbf{q}) A_x(\mathbf{q}),
\end{equation}
where $A_x(\mathbf{q})=iB(\mathbf{q})/q_y$ is the vector potential 
in an appropriate gauge.

Extrapolating
$K_{xx}
$ to the thermodynamic limit 
is challenging 
due to the presence of strong finite--size effects associated with the Fermi surface. These finite size effects are dramatically reduced by performing the computations 
in the presence of a weak a uniform magnetic field~\cite{Assaad2002a}. We adopt this procedure, but make the field spin-dependent, with the total flux through the system being $2\pi$ ($-2\pi$) for spin up (spin down).
As shown in Ref.~\cite{Wu2005}, such a procedure does not introduce a sign problem for the Monte Carlo. 
Moreover, in the limit of large system size, the magnetic field vanishes. (See Appendix~\ref{app:DQMC} for the implementation of this effective magnetic field in the simulations.)

Fig.~\ref{fig:rho_s} shows
$K_{xx}$ 
as a function of $q_y$, for different 
system sizes at temperature $T=0.05$ and $h=h_c$.
The data 
 collapse onto a single curve 
 that extrapolates to a value 
 near zero in the limit $q\rightarrow 0$. 
 The extrapolated value is clearly much below the universal value at the BKT transition, $\rho^{BKT}_s = 2T/\pi$. This 
 implies that at $h = h_c$, the system is not superconducting down to $T = 0.05t$. Similarly, the superfluid density for $h \ne h_c$ is found to be below $\rho^{BKT}$ for all $T\ge 0.05t$. Note that, from Eq.~(\ref{eq:Kxx}), $\mathrm{lim}_{q_y \rightarrow 0} K_{xx}/q_y^2$ is the 
 orbital susceptibility. From Fig.~\ref{fig:rho_s}, we see that in our system, this quantity is negative, hence the orbital response is paramagnetic~\cite{comment_diamagnetic}. This indicates that there are no strong superconducting fluctuations (which 
 would produce diamagnetism).

We have also computed the superconducting susceptibilities in the $s$--wave ($A_{1g}$) and $d$--wave (B$_{1g}$) channels, as a function of $h$ and $T$:
\begin{equation}
P_{s,d} = \int_0^\beta d\tau \sum_i \langle \Delta_{s,d}^\dagger (\mathbf{r}_i, \tau)  \Delta^{\vphantom{\dagger}}_{s,d}(0,0) \rangle.
\label{eq:sc_suscept}
\end{equation}
Here, $\Delta_s(\mathbf{r}_i) = c_{i\uparrow} c_{i\downarrow}$ and $\Delta_d(\mathbf{r}_i) = \sum_j \eta_{ij} (c_{i\uparrow} c_{j\downarrow} - c_{i\downarrow} c_{j\uparrow})$, where $\eta_{ij}$ is defined below Eq.~(\ref{eq:chi1}).
The superconducting susceptibilities are displayed in Fig.~\ref{fig:P_of_T}. As expected for a non-superconducting state, both $P_s$ and $P_d$ saturate as a function of system size at fixed $T$ 
and $h$ [see Fig.~\ref{fig:P_of_T}(a,b)]. $P_s$ has a 
maximum at $h \approx h_c$, while $P_d$ is 
smaller and more or less independent of $h$. 

The enhancement of $P_s$ by nematic fluctuations is 
further reflected in its behavior as a function of temperature at a fixed $h$ [Fig.~\ref{fig:P_of_T}(c,d)].
$P_s$ rises more rapidly than logarithmically at low temperatures.  In contrast for non-interacting electrons 
$P \sim 
\log (T_0/T)$, 
a functional form consistent with the $T$ dependence of $P_d$ at low $T$. We interpret this as an indication that the $s$--wave susceptibility is enhanced by induced attractive interactions mediated by nematic fluctuations; the ground state for $h$ near $h_c$ is likely 
 an $s$--wave superconductor, although with $T_c < 0.05t$.

\subsection{Single Fermion Correlations}
\label{sec:fermionic}

{Having found that the system is not superconducting
down to the temperatures reached in this study, we examine the single-fermion Green's function to glean information about the metallic state in the vicinity of the QCP. 
The metal is naturally characterized via the single particle density of states $N(\omega)$, the Fermi velocity ${\bf v_F}$, and the quasiparticle weight $Z_{{\bf k_F}}$. However, measuring these quantities within DQMC is not straightforward, since they generally require real-frequency information inaccessible without analytic continuation. In addition, ${\bf v_F}$ and $Z_{{\bf k_F}}$ are well defined only at zero temperature, while our simulations are at finite temperature. We partially resolve these difficulties by defining finite temperature objects $\widetilde N(T)$, ${\bf v_F}(T)$, and $Z_{\bf k_F}(T)$ which can be measured numerically, and whose zero temperature limits are $N(\omega=0)$, ${\bf v_F}$, and $Z_{{\bf k_F}}$ respectively. Readers interested in the raw frequency dependence of the fermionic Green's function and self-energy are referred to appendix \ref{app:G_and_sigma}.

Real frequency information can be extracted from our imaginary-time data using the identity~\cite{Trivedi1995}:
 \begin{equation}
G(\mathbf{k},\tau>0) = \int_{-\infty}^\infty d\omega \frac{e^{-\omega(\tau-\beta/2)}}{2\cosh({\beta \omega }/2)}A(\mathbf{k}, \omega),
\label{eq:G2}
\end{equation}
where $A({\bf k},\omega)$ is the fermion spectral function. $G({\bf k},\beta/2)$ is evidently equal to the spectral function integrated over a range of energies $\sim T$.  We therefore define

 \begin{align}
\widetilde{N}(T) \equiv & \frac{1}{T}\int \frac{d^2k}{(2\pi)^2} G(\mathbf{k},\tau=\beta/2)
\nonumber \\
=& \int_{-\infty}^{\infty} \frac{ d\omega}{2 T \cosh(\beta \omega/2)} N(\omega),
\label{eq:Ntilde}
\end{align}

\begin{align}
 {\bf v_F}(T)\equiv \frac{\partial^2}{\partial{\bf k}\partial\tau} \log\left[G({\bf k},\tau)\right]  |_{{\bf k}={\bf k_F},\tau=\beta/2},
 \label{vf}
\end{align}

\begin{align}
Z_{{\bf k_F}}(T)\equiv 2 G({\bf k_F},\tau=\beta/2).
\label{eq:Zdef}
\end{align}
We explain in Appendix~\ref{app:metallic} how these quantities attain the desired zero temperature limits. We also discuss in the appendix the procedures used to estimate these quantities in finite size systems with discrete momentum and imaginary time.}

\begin{figure}[t]
\centering
\includegraphics[width=0.7\columnwidth]{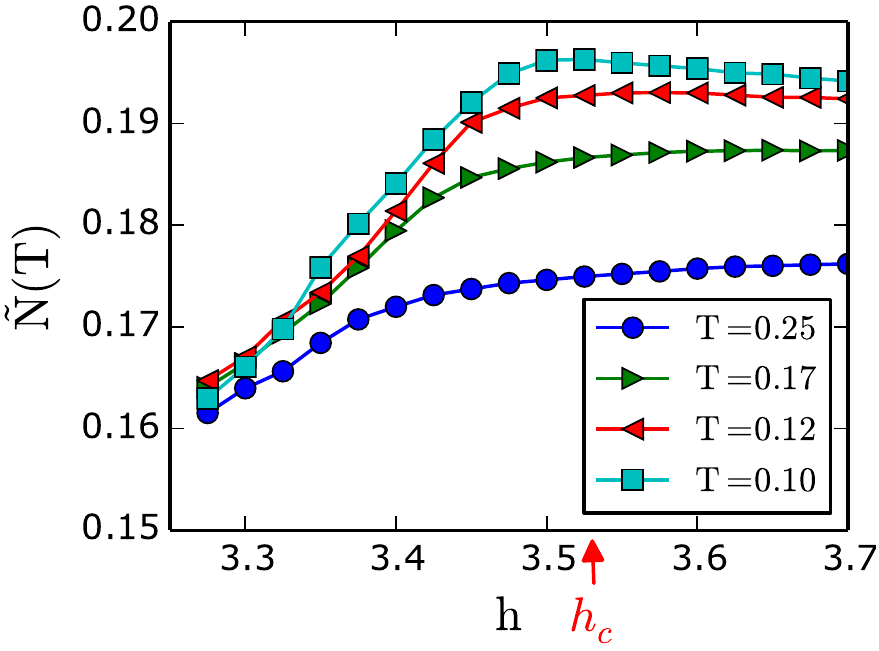}
\caption{ $\tilde{N}(T)$, defined in Eq.~(\ref{eq:Ntilde}), as a function of $h$ 
 for  system  size $L=20$ at different temperatures. 
 $\tilde{N}$ equals 
 the density of states at the Fermi energy in the limit $T\rightarrow 0$.  }
\label{fig:DOS}
\end{figure}

Fig.~\ref{fig:DOS} shows $\widetilde{N}(T)$ for an $L=20$ system as a function of $h$, for different temperatures. 
$\widetilde{N}(T)$ 
remains nonzero for all values of $h$, consistent 
 with a finite density of states in the limit $T\rightarrow 0$.  
 This indicates 
 the absence of a substantial gap 
  in the fermionic spectrum, even a 
  gap with nodal points on the Fermi surface, down to $T=t/12$, 
corroborating the conclusion of Sec.~\ref{sec:SC} that the system is not superconducting down to these temperatures. 
Moreover, the fact that $\widetilde{N}(T)$ 
seems to neither vanish nor diverge at low temperature and $h = h_c$ has important implications for the nature of the metallic state of the QCP.
Were the fermions to acquire a positive (negative) anomalous dimension, $\eta_F$,
the density of states would go to zero (diverge) at $T=0$~\cite{Metlitski2010,Senthil2008}. We elaborate on this point in Sec.~\ref{sec:discussion}. Instead, the saturation of the density of states is consistent with a Fermi liquid, at least down to the temperature scales in our simulations. Note, as well, that the density of states appears to be depressed with increasing nematic order for $h<h_c$

\begin{figure}[t]
\centering
\includegraphics[width=1.0\columnwidth]{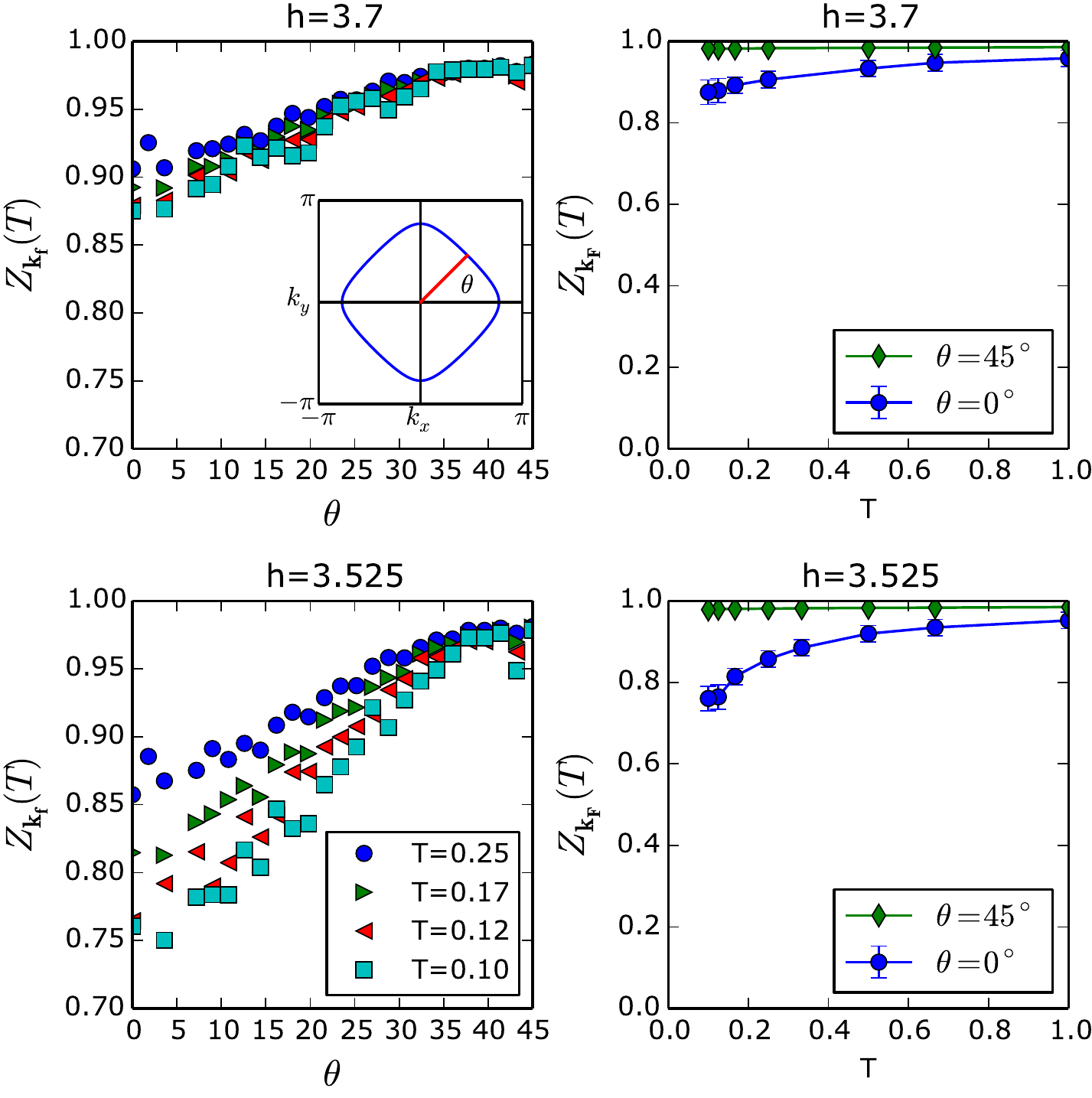}
\caption{Left panels 
show $Z_{\mathbf{k_F}}$ [defined in Eq.~\ref{eq:Zdef}] as a function of angle $\theta$ along the Fermi surface, for different values of $h$. The QCP is at $h_c \approx 3.525$. Right panels show the temperature dependence of $Z_{\bf k_F}(T)$ for high symmetry directions of $k_F$.
$Z_{\mathbf{k_F}}$ approaches the conventionally defined quasiparticle weight as $T\to 0$.}
\label{fig:Z}
\end{figure}

\begin{figure}[t]
\centering
\includegraphics[width=1.0\columnwidth]{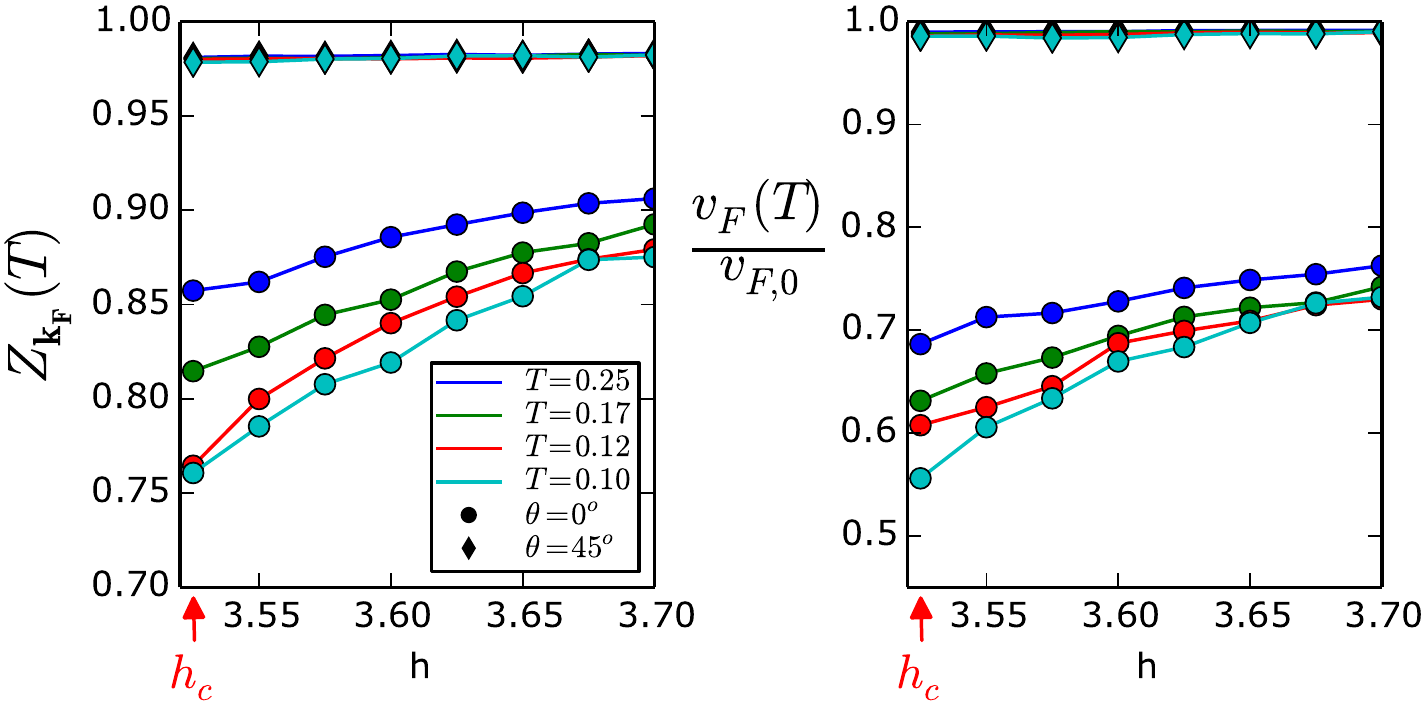}
\caption{(a) 
$Z_{\mathbf{k_F}}$ as a function of $h$ for different temperatures. Circles (diamonds) correspond to $\theta = 45^\circ$ ($\theta = 0^\circ$), respectively (where $\theta$ is defined in the inset of Fig.~\ref{fig:Z}). (b) The Fermi velocity renornalization, $v_F / v_{F,0}$ (where
$v_F$ is defined in Eq.~\ref{vf} and $v_{F,0} $ is the non-interacting Fermi velocity) as a function of $h$ for $\theta = 0^\circ$, $45^\circ$.}
\label{fig:vf}
\end{figure}

The results for $Z_{\mathbf{k_F}}(\theta,T) $, where $\theta$ is the angle between $\mathbf{k_F}$ and the $x$ axis, are shown in Fig.~\ref{fig:Z} for different values of $h \ge h_c$. At $\theta = 45^{\circ}$, $Z_{\mathbf{k_F}}$ is close to unity at all temperatures. This can be understood as a consequence of the fact that at $\theta = 45^{\circ}$, the coupling between fermions and the nematic critical modes vanishes by symmetry; hence, at these ``cold spots'', the effects of scattering off the critical modes are minimal. $Z_{\mathbf{k_F}}$ decreases systematically as $\theta$ approaches $0^\circ$, and as $h$ approaches $h_c$. $Z(k_F,\theta=0,T)$ decreases with decreasing temperature with negative curvature. On the basis of the present data, it is difficult to say whether it extrapolates to a finite value in the $T\rightarrow 0$ limit (indicating a FL ground state), or vanishes with a 
 small exponent, $Z\sim T^{a_Z}$ where $a_Z \approx 0.1$ (indicating 
a breakdown of FL behavior). It would be difficult to reconcile the data with 
 $a_Z$ larger than about $0.15$.

Fig.~\ref{fig:vf}(a) shows the same data as a function of $h \ge h_c$ for $\theta = 0^\circ$ and $\theta=45^\circ$, emphasizing the decrease in $Z_{\mathbf{k_F}}$ as $h$ approaches $h_c$. Measuring $Z_{\mathbf{k_F}}$ close to the QCP for $h<h_c$ (in the nematic phase) has proven difficult, since the two configurations of the order parameter correspond to two different Fermi surfaces with a small splitting between them. In Fig.~\ref{fig:vf}(b) we show the renomalization of the Fermi velocity, $v_F / v_{F,0}$, where $v_{F,0} $ is the Fermi velocity for the non-interacting band structure ($\alpha = 0$). As with $Z_{\mathbf{k_F}}$, $v_F$ is hardly renormalized at the cold spot $\theta = 45^\circ$. In contrast, at $\theta = 0^\circ$, $v_F$ is strongly reduced from its non-interacting value, and the velocity renormalization increases as $h$ approaches $h_c$. However, down to $T = 0.1t$, there is no obvious indication 
that $v_F$ vanishes (i.e., the effective mass diverges) as $T \rightarrow 0$, even at $h = h_c$.

Over the temperature range explored in this study, the single fermion Green's function
is consistent with a renormalized Fermi liquid, with an angle- and $h$- dependent quasi-particle weight and velocity. 
However, the expected differences between a Fermi liquid and a ``marginal Fermi liquid'' or a weak (small $a_Z$) non Fermi liquid are subtle at non-zero $T$.
We can 
 rule out a non-Fermi liquid with a substantial value of $a_Z$ and/or $\eta_F$, but not one with $a_z$ and $\eta_F$ sufficiently small.

\section{Discussion}
\label{sec:discussion}

We view our results as a numerical ``experiment.''  For an explicit lattice Hamiltonian, we have obtained detailed and extensive ``measurements'' of physical quantities.  In the neighborhood of the quantum phase transition we find an enhanced tendency to superconductivity, substantial scattering of the fermionic quasiparticles, and a clearly delineated quantum  critical fan.  Moreover, for  a large range of parameters, a simple scaling form  provides a good description of the thermodynamic nematic fluctuations.
In laboratory experiments on solid state systems, good scaling collapse over a similar dynamical range is often interpreted in terms of universal critical exponents.
However, in our system, the fermion bilinear quadrupolar correlations are only consistent with the apparent scaling form of the nematic correlator over a narrow region as a function of $(T, h-h_c, \mathbf{q})$. This is reason to question whether the behavior we are seeing is characteristic of the asymptotic approach to a quantum critical point.
It is possible that the success of the scaling analysis is fortuitous.  Alternatively, the observed scaling may reflect an intermediate asymptotic regime, possibly associated with an unstable multicritical point.  It is also possible that  corrections to universal scaling behavior are simply smaller for $D$ than for $Q$, in which case the scaling form of $D$ might reflect the true critical behavior.    At the very least, the observation of  simple power laws invites a comparison with  theories of metallic quantum criticality.

\subsection{Comparison with 
 theory}
The theory of metallic QCPs, along with the closely related theory of metals coupled to fluctuating gauge fields\cite{Holstein1973,Reizer1989,Halperin1993}, has traditionally been treated by integrating out the fermions to obtain an effective action for the bosonic modes that is highly non-local in both space and time.  In this ``Hertz-Millis'' approach, a renormalization group (RG) analysis is then undertaken for the effective field theory so defined.  While much can be learned
from this program, there are at least two general causes for concern:  1) The analyticity of the $\beta$-function, which is a core ingredient of  RG, is typically ensured by the locality of the action. For a non-local action, all aspects of the analysis, especially the enumeration of potentially relevant interactions that can be generated under RG, need to be tested explicitly.  2)  In many cases, including in particular the 2D nematic QCP~\cite{oganesyan-2001,Metzner2003}, if one considers the back effect of the collective fluctuations on the fermions, one finds that Fermi liquid theory is perturbatively unstable. 
The question then arises concerning the self-consistency of the effective action which is obtained by integrating out an assumed Fermi liquid.

Nonetheless, there is a general feeling that this approach is 
reliable in large enough spatial dimension~\cite{sachdev2007quantum}
\footnote{
In $d=0$ as well, the Kondo problem can be mapped\cite{Chakravarty1984} to the problem of a two-level system coupled to an Ohmic heat bath, for which the non-local effective action is equivalent to that of a $d=1$ classical inverse-square Ising model. This problem can, in turn, be treated using a suitable RG analysis~~\cite{anderson1970exact}.}, and its main results are similar in structure to those of many subsequent studies. Accordingly, we briefly review the main features of the Hertz-Millis treatment of the Ising nematic QCP.  

In this context, the quadratic part of the effective action for the nematic mode is of the form
\begin{equation}
D_0^{-1}(\mathbf{q},\omega_n) = r + \kappa |\mathbf{q}|^2 + \alpha^2 Q_0(\mathbf{q},\omega_n)
\label{HM}
\end{equation}
where $r$ and $\kappa$ are  (renormalized) functions of $T$ and $h$, and $Q_0$ is the suitably defined particle-hole response function (with the zero-frequency piece subtracted)

\begin{equation}
Q_0(\mathbf{q},\omega_n) \equiv \int \frac {d\mathbf{k}}{(2\pi)^2}\frac {i\omega_n|g(\mathbf{k},\mathbf{q})|^2 [f(\varepsilon_{\mathbf{k}+\mathbf{q}}) -f(\varepsilon_{\mathbf{k}})]}
{[\varepsilon_{\mathbf{k}+\mathbf{q}}-\varepsilon_{\mathbf{k}}][i\omega_n+\varepsilon_{\mathbf{k}+\mathbf{q}}-\varepsilon_{\mathbf{k}}]},
\label{Q}
\end{equation}
{where $g(\mathbf{k},\mathbf{q})$ is a dimensionless form factor of appropriate symmetry, 
 whose 
  form depends on microscopic details (
  e.g.   $g(\mathbf{k},\mathbf{q}) =\cos(k_x)-\cos(k_y)$).}
The essential features of this expression relate to its asymptotic (IR) behavior in the limit $E_F \gg
{ |\mathbf{v_F}(\mathbf{\hat{k}_q})\times \mathbf{q}|}\gg |\omega_n|$, 
\begin{equation}
\label{Q_asymptotics}
Q_0(\mathbf{q},\omega_n) \sim |g(\hat{\mathbf{k}}_{\mathbf{q}},0)|^2\frac{|\omega_n|}{
{ |\mathbf{v_F}(\mathbf{\hat{k}_q})\times \mathbf{q}|}},
\end{equation}
where $\hat{\mathbf{k}}_{\mathbf{q}}$ is the point on the Fermi surface (assumed to be unique mod inversion) at which the Fermi velocity, $\mathbf{v_F}(\mathbf{\hat{k}_q})$, is perpendicular to ${\mathbf{q}}$.  (The fact that the integral in Eq. (\ref{Q})
for given $\mathbf{q}$ is dominated by the neighborhood of this point on the Fermi surface is the essential observation motivating the ``patch'' theories of this problem.)

The Hertz-Millis approach implies that the dynamical exponent $z=3$ at tree level. From this, a naive scaling analysis (neglecting the possibility of higher-order non-local interactions) concludes that, since $d+z=5$ is well above the Ising upper critical dimension,  mean-field exponents $\nu=1/2$, $\gamma=1$, and $\eta=0$ apply,
consistent with what we find.  (See Table I.)  Moreover, one would expect violations of scaling (for instance $\tilde z\neq z$ and 
 $\lambda\neq 1/\nu z$). 
 Indeed in Refs.~\cite{Millis1993,Jakubczyk2008,Bauer2011,Hartnoll2014} it is shown that at criticality, such an 
 analysis leads to $r\sim T\log[E_F/T]$.  As we do not have the dynamical range to detect a logarithm, this, too, is consistent with our results.

However, on the face of it, our results for the dynamical nematic response are inconsistent with a dynamical exponent $z=3$ and with the sort of anisotropies one might expect from the patch construction.
 Moreover, the relative weakness of any deviations of the single-particle properties from Fermi liquid behavior appears inconsistent with the Hertz-Millis predictions~\footnote{To our knowledge, the temperature dependence of the quantity $Z(T)$ defined here has not been explicitly computed within the Hertz-Millis approach. Applying $\omega/T$ scaling to the $T=0$ self energy (which, of course may not be valid) yields $Z(T)\sim T^{1/3}$, a noticeably more rapid variation than the data of Fig. \ref{fig:Z}d.}.
On the other hand, at the smallest momenta, deviations 
of $D$ from the scaling form, ${\cal A}$, Eq.~(\ref{eq:D_dynamics}), become apparent. This may indicate a crossover \footnote{Though our Fermi surface is large, there is a small spanning vector $Q=(2\pi) \times 0.23$ connecting Fermi surface copies in adjacent Brillouin zones (see Fig.~\ref{fig:fermi_surface}), and this might lead to  a crossover scale at small momentum.} from an intermediate $z\approx 2$ regime to a different behavior
in the deep infra-red~\cite{note-preliminary}.

The foregoing discussion of 
the relation between our data and Hertz-Millis theory can be applied to numerous subsequent attempts to bring the problem under theoretical control. Among the approaches employed by these studies, whose results have similar structure to those of Hertz-Millis, are large $N_F$ methods (where $N_F$ is the number of fermion flavors)\cite{altshuler1994low,Altshuler1995,Abanov1999}, dimensional\cite{Chakravarty1995, Senthil2009, Dalidovich2013} and dynamical\cite{nayak1994non,nayak1994renormalization,mross2010controlled} regularization, and higher dimensional bosonization\cite{lawler-2006,lawler-2007}. The related problem of orbital loop current criticality in a metal has been argued\cite{Varma1997,Varma1999} to lead to a critical bosonic {correlator} with $|\omega|$ frequency dependence (as seen in our results), but with an asymptotic absence of momentum dependence (not seen in our results) characteristic of ``local quantum criticality"~\cite{si2001locally}. In the vicinity of $d=3$, several varieties of intermediate asymptotics corresponding to different schemes of large $N_F$ and $N_B$ (the number of boson flavors)  have also been explored\cite{fitzpatrick1,fitzpatrick}, at least one of which has $z=2$ upon naive extrapolation to $d=2$. {We believe our results motivate revisiting  existing theories of nematic quantum criticality in metals. 
In particular, the isotropy of bosonic correlations we observe 
appears to be at odds with ``patch" constructions.}

Finally, there has been considerable theoretical interest in the issue of superconductivity in the neighborhood of a nematic quantum critical point.  It has been argued \cite{Abanov2001Incoherent,roussev2001quantum,metlitski2010instabilities,metlitski2015cooper} that superconductivity with a critical temperature of order 1 times microscopic scales should be expected 
near the nematic quantum critical point.  Indeed, it has been suggested that the superconducting $T_c$ at criticality may be so high that it preempts the quantum critical regime in which non-Fermi liquid behavior would be expected.  More modestly, in the small $\alpha$ limit, it has been shown\cite{lederer2015enhancement} that nematic fluctuations do indeed mediate attractive pairing interactions, and that the resulting transition temperature grows singularly as the quantum critical point is approached either from the ordered or the disordered side.  In agreement with this latter result, the superconducting susceptibilities shown in  Fig. \ref{fig:P_of_T}  do show a maximum at $h=h_c$.  However, if $T_c$ at criticality is a number  of order 1 times the Fermi energy, that number is apparently sufficiently small that $T_c < E_F/70$ for our chosen parameters.  {At larger values of the boson-fermion coupling constant, preliminary results~\cite{Lederer-future} show that the ground state is an $s-$wave superconductor whose $T_c$ is maximal near the nematic QCP.}

\subsection{Relation to experiment}

There is increasingly extensive evidence that electron nematic phases are common in highly correlated electronic fluids.\cite{Fradkin2010}  In particular, it has been suggested that a nematic QCP occurs near optimal doping in both the cuprates\cite{kivelson1998electronic,nie2014quenched} and in the Fe-based superconductors\cite{fang2008theory,xu2008ising}.  However, the experimental situation in all these cases is complicated, and in some cases controversial.

It is premature to attempt any sort of serious comparison between the present results and the experiments  in these materials.  However, the experimental case is clearest in certain Fe based superconductors, so it is worth noting a few points of comparison.  Most strikingly, elastoresistance~\cite{Chu2012,KouFisher2015}, Raman scattering~\cite{Gallais2013,  Blumberg2014}, and elastic constant~\cite{Bohmer2014} measurements show a large range of $T$ and doping over which a large nematic susceptibility can be documented with a remarkably systematic Currie Weiss $T$ dependence $\chi \sim 1/[T-T^*(x)]$ where $T^*(x)$ appears to depend roughly linearly on ``doping'' concentration, $x$, and to pass through zero at a critical doping concentration, $x_c$, which approximately coincides with the optimal doping for superconductivity.
It is impossible not to be encouraged by the similarity between this experimental finding and our numerical results.

{Note added:  After our paper was posted, a new theoretical analysis of the finite $T$ dynamics near a nematic metallic QCP~\cite{Punk} has found a possible route to understanding the observed behavior of $D$, and in particular for the apparent dynamical exponent $z=2$. }

\acknowledgements{We are grateful to D. Chowdhury, A. Chubukov, S. Hartnoll, R. Fernandes, E. Fradkin, A. Millis, S. Raghu, and S. Sachdev for illuminating discussions. E. B. and Y. S. were supported by the ISF under grant 1291/12, by the US-Israel BSF, and by a Marie Curie CIG grant.
SAK was supported in part by NSF grant \#DMR 1265593 at Stanford.  SL was supported, in part by DOE, grant \# DE-AC02-76SF00515,
an ABB Fellowship at Stanford, and a Gordon and Betty Moore Post Doctoral Fellowship at MIT.}

\appendix

\section{Technical details of the DQMC simulations}
\label{app:DQMC}
Discretizing imaginary time as $\beta=N\Delta\tau$, the grand partition function becomes
\begin{equation}
\Xi=\mathrm{Tr}_{\lbrace\tau\rbrace}\mathrm{Tr}_{\lbrace c,c^{\dagger}\rbrace}\prod_{n=1}^{N/2}\left(\hat{B}\hat{B^{\dagger}}\right)+O(\Delta\tau^{2}),
\end{equation}
where
\begin{equation}
\hat{B}=e^{-\frac{\Delta\tau}{2}H_{h}}e^{-\Delta\tau H_{V}}e^{-\Delta\tau H_{\mu}}\prod_{m=1}^{4}\left(e^{-\Delta\tau c^{\dagger}K^{(m)}c}\right)e^{-\frac{\Delta\tau}{2}H_{h}}.
\end{equation}
Here, $c^{\dagger} = (c_{1,\uparrow}^{\dagger},c_{1,\downarrow}^{\dagger},c_{2,\uparrow}^{\dagger},c_{2,\downarrow}^{\dagger},...)$ is a row vector of spin $1/2$ fermionic creation operators for the spatial sites $1,2,...,L^2$, where $L$ is the linear dimension. $H_h,H_V,$ and $H_{\mu}$ are the terms in Eq. \ref{eq:HFHB} proportional to $h,V,$ and $\mu$ respectively. We use a checkerboard decomposition to describe the kinetic energy matrices $K^{(m)}$, whose elements are $K_{i,j,\sigma,\sigma'}^{(m)}=-t_{i,j}^{(m)}(\sigma)(1-\alpha\tau_{i,j}^{z})\delta_{\sigma,\sigma'}$, i.e, $m=1,2,3,4$ enumerates the horizontal/vertical bonds originating from a site with an even/odd index. We allow $t_{i,j}$ to depend on spin in order to implement the spin-dependent magnetic field discussed below.

Plugging in unity operators in the $\tau_{ij}$ sector at every time slice and taking the trace over the fermions, we bring the partition function to a form which can be sampled using Monte-Carlo techniques:
\begin{equation}
\Xi=\sum_{\lbrace\tau_{i,j;n}=\pm1\rbrace}e^{-S_{\tau}}\det\left(1+e^{\beta\mu}\prod_{n=1}^{N/2}T_{2n-1}T_{2n}^{\dagger}\right)+O(\Delta\tau^{2}),
\label{eq:Z}
\end{equation}
where $T_{n}=\prod_{m=1}^{4}\left(e^{-\Delta\tau K_{n}^{(m)}}\right)$ is a matrix which depends implicitly on the c-numbers $\tau_{i,j;n}=\pm 1$ through the relation $K_{i,j,\sigma,\sigma';n}^{(m)}=-t_{i,j}^{(m)}(\sigma)(1-\alpha\tau_{i,j;n})\delta_{\sigma,\sigma'}$, and the bosonic part of the action is given by
\begin{eqnarray}
S_{\tau} &=& \frac{1}{2}\log\left(\tanh(\Delta\tau h)\right)\sum_{\langle i,j\rangle,n}\tau_{i,j;n}\tau_{i,j;n+1} \nonumber \\ &+& V\Delta\tau\sum_{\langle\langle i,j;k,l\rangle\rangle,n}\tau_{i,j;n}\tau_{k,l;n}.
\label{eq:S_tau}
\end{eqnarray}
We have used $\Delta \tau = 0.05$ in our simulations. 

Monte-Carlo sampling is most efficient when the determinant in \eqref{eq:Z} is positive. Here, the absence of a sign problem is be guaranteed by microscopic time reversal symmetry. To see this, note that the matrices $T_n$ are block diagonal in the spin sector, and so the total determinant is the product of the spin up and spin down determinants. 
If time reversal is preserved
, i.e
\begin{equation}
t_{i,j} (\uparrow) = t^*_{i,j} (\downarrow),
\label{eq:TRS}
\end{equation}
then the product of the determinants is clearly positive.

It has long been known\cite{Assaad2002a} that finite size effects in fermionic systems, in particular at low temperatures, can be greatly improved by the application of a weak orbital magnetic field. On a finite size system with periodic boundary conditions, the field must be such that an integer multiple of the flux quantum passes through the system, $\Phi = n \Phi_0$. In order to 
preserve time reversal symmetry as required in \eqref{eq:TRS}, in this work we have applied the opposite field for both spin species, such that $\Phi_\uparrow  = -\Phi_\downarrow= \Phi_0$. 
Such a field vanishes in the thermodynamic limit.

Although our simulations do not suffer from the sign problem, close to the quantum critical point our simulations do suffer from critical slowing down. To avert this, we found it was helpful to include global Monte-Carlo updates, in addition to local (Metropolis) ones. Our global updates consisted of a slight modification to the Wolff algorithm\cite{Wolff1989}: Starting from a space-time configuration of pseudo-spins, a cluster of pseudo-spins is constructed using the usual Wolff algorithm, supposing our action is given by \eqref{eq:S_tau}, i.e without taking into account the coupling to the fermions. Next, we propose flipping the pseudo-spins which belong to the cluster, and accept the move with a probability which is the ratio of the determinants in \eqref{eq:Z} between the two configurations. It can easily be seen that such a move obeys detailed balance.
\section{Finite size scaling analysis}
\label{app:finite}
\begin{figure}
\includegraphics[width=1.0\columnwidth]{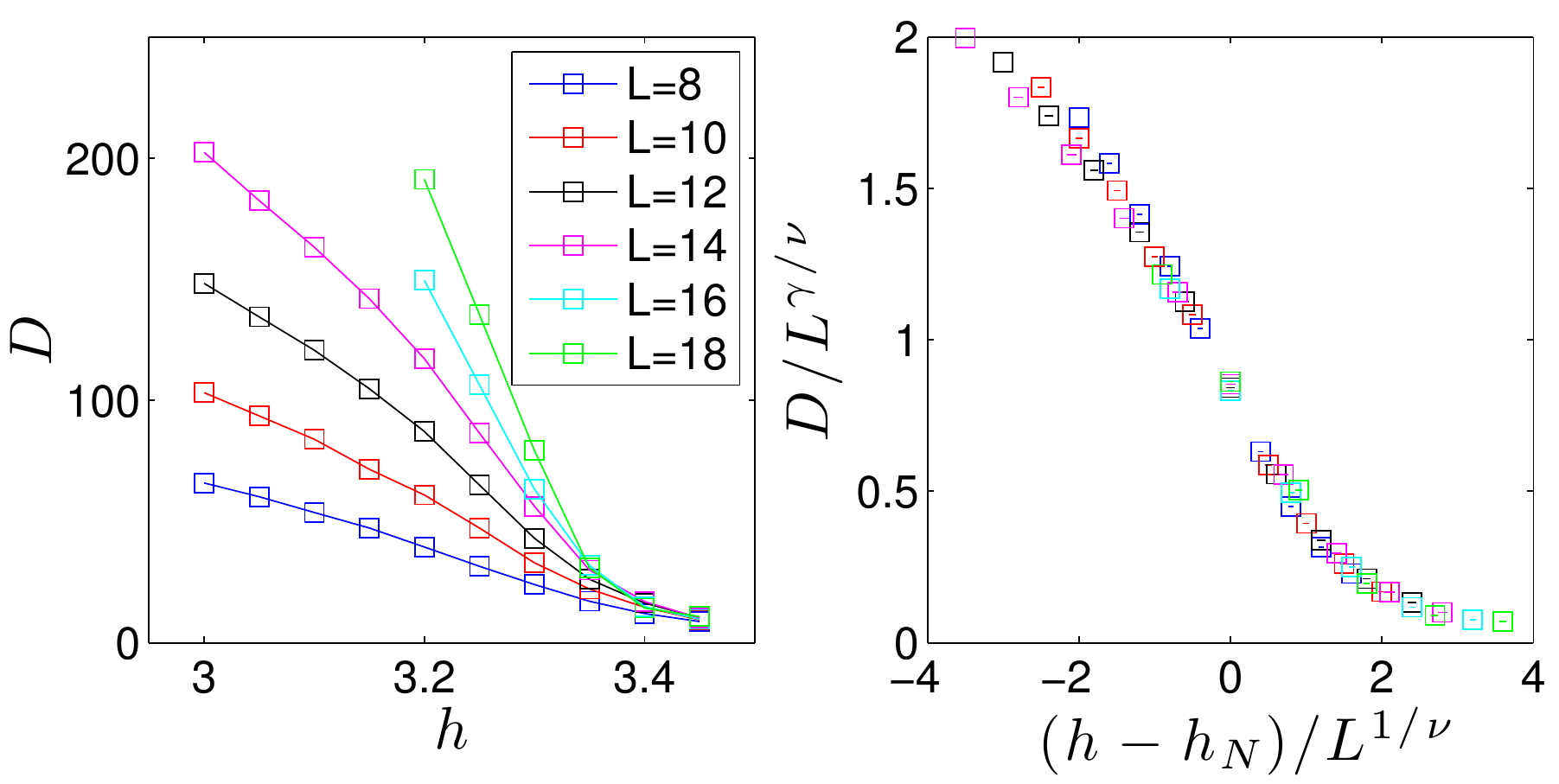}
\includegraphics[width=1.0\columnwidth]{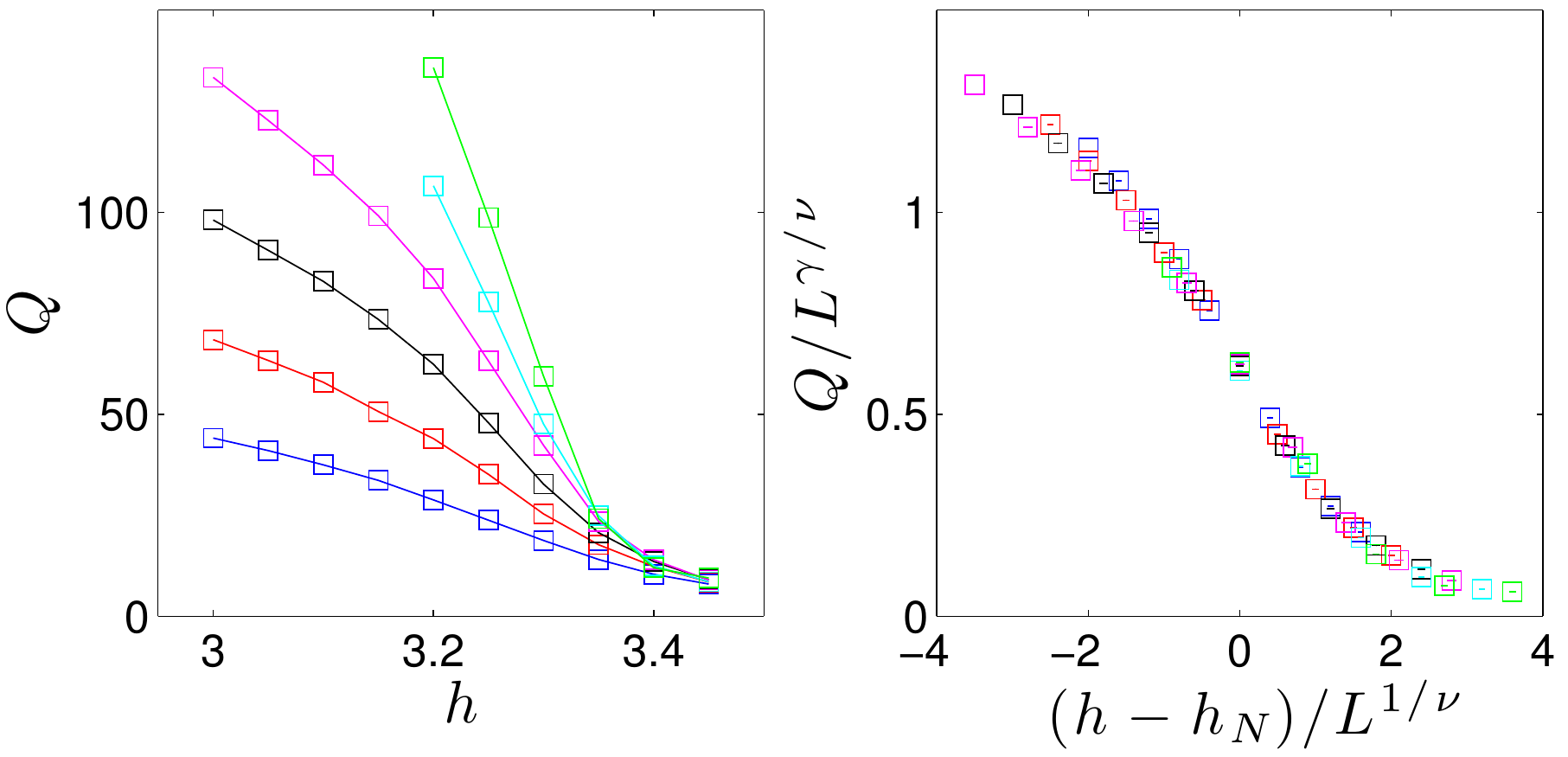}
\caption{An example of the procedure used to calculate the critical value of $h$ for a given temperature. Here, for $T=0.33$, the best data collapse is obtained for $h_N\approx 3.25$. The exponents take the two dimensional Ising values $\gamma=7/4$,$\nu=1$ for this thermal transition. The nematic susceptibility and quadrupolar susceptibility have similar behavior.}
\label{fig:classical_collapse}
\end{figure}

The finite temperature phase boundary $T_N(h)$ (or equivalently $h_N(T)$) can be computed by using {\it classical} finite size scaling techniques for two dimensional Ising transitions, characterized by susceptibility exponent $\gamma=7/4$ and and correlation length exponent $\nu=1$. At a given temperature and for a given system size $L$, the most singular part of the thermodynamic susceptibility, $\chi_s$, satisfies
\begin{equation}
\chi_s(h;L)=L^{\gamma/\nu}F((h-h_N)L^{1/\nu})
\end{equation}
$h_N$ can therefore be identified using a single parameter data collapse, as illustrated in Fig. \ref{fig:classical_collapse}. At low temperatures, the region of parameter space corresponding to classical criticality narrows, so that larger system sizes are needed to reliably estimate $h_N$.

\begin{figure}[t]
\centering
\includegraphics[width=1.0\columnwidth]{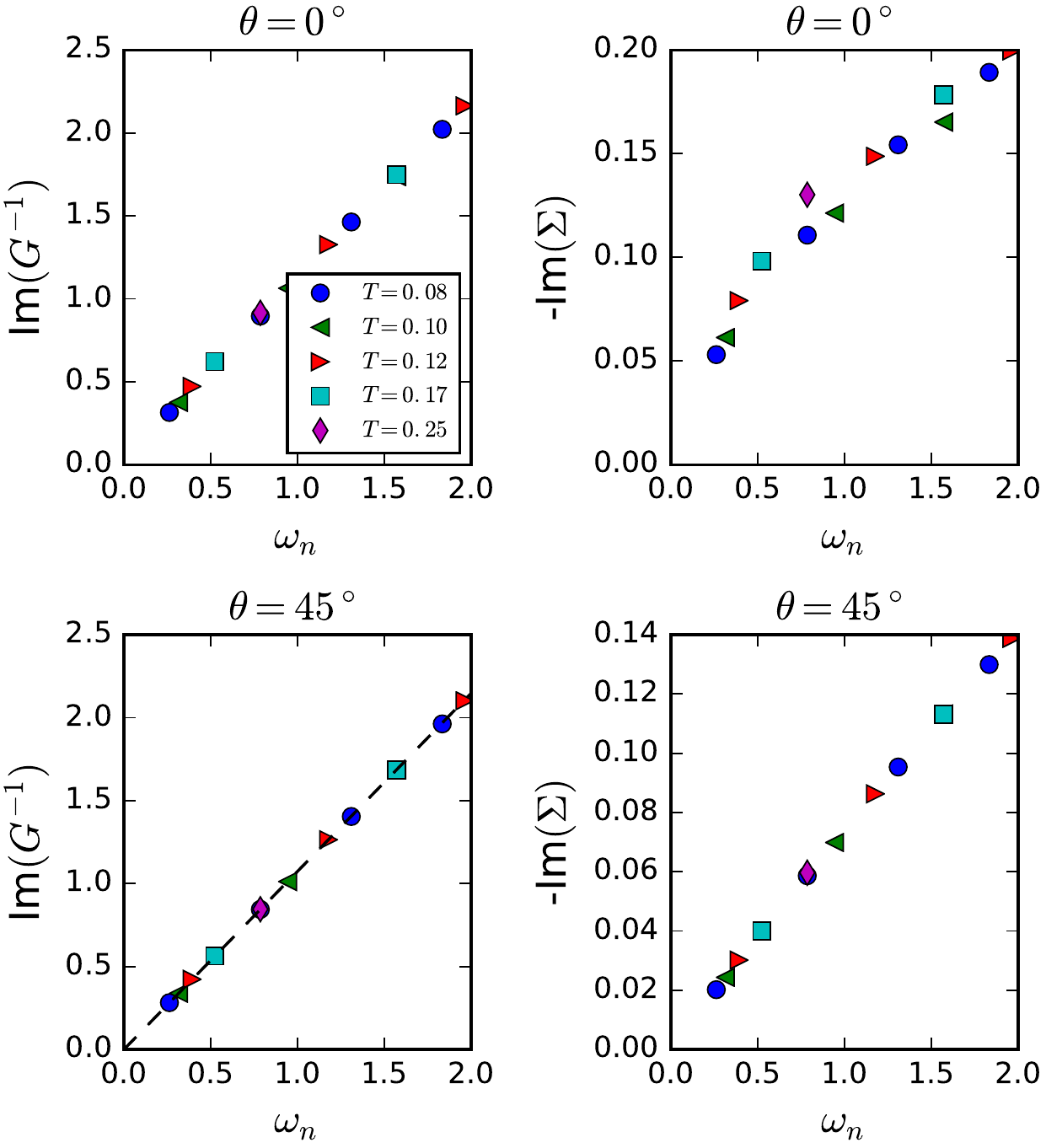}
\caption{Left panels: The imaginary part of the inverse of the Matsubara Green's function $\mathrm{Im}(G^{-1}(\mathbf{k_F}, \omega_n))$ for $h=3.525$. Data points from different temperatures mostly lie on the same curve. Right panels: The imaginary part of the self energy. The dashed line in the bottom left panel corresponds to $Z=0.93$.}
\label{fig:matsubara_G}
\end{figure}

\section{Characterizing the metallic state}
\label{app:metallic}
 Write the imaginary time Green's function in a spectral representation:
\begin{eqnarray}
G(\mathbf{k},\tau>0) &=& \langle  c_{\mathbf{k}}^{\vphantom{\dagger}}(\tau) c^\dagger_{\mathbf{k}}(0) \rangle \nonumber \\
&=&\sum_{m,n} \frac{e^{-\beta \varepsilon_n}}{\Xi} e^{\tau (\varepsilon_n - \varepsilon_m )} |\langle n \vert c_{\mathbf{k}} \vert m \rangle |^2.
\label{eq:G1}
\end{eqnarray}
Here, $\vert n \rangle$ is a many-body eigenstate with energy $\varepsilon_n$, and $\Xi$ is the grand partition function. 
In this representation, the fermion spectral function is
\begin{equation}
A(\mathbf{k}, \omega) = \sum_{m,n} \frac{e^{- \beta \varepsilon_n} + e^{- \beta \varepsilon_m }}{\Xi} |\langle n \vert c_{\mathbf{k}} \vert m \rangle |^2 \delta(\omega + \varepsilon_n - \varepsilon_m),
\end{equation}
we can write Eq.~(\ref{eq:G1}) as~\cite{Trivedi1995}
\begin{equation}
G(\mathbf{k},\tau>0) = \int_{-\infty}^\infty d\omega \frac{e^{-\omega (\tau-\frac{\beta}{2})} }{2\cosh({\beta \omega }/2)} A(\mathbf{k}, \omega).
\label{eq:G3}
\end{equation}

If the ground state is a Fermi liquid, we can extract the quasi-particle weight and Fermi velocity renormalization from Eq.~(\ref{eq:G3}). We assume that at low temperatures, the spectral function in the vicinity of the Fermi surface obtains a Fermi liquid form, $A(\mathbf{k},\omega) = A_{\mathrm{qp}}(\mathbf{k},\omega) + A_{\mathrm{reg}}(\mathbf{k},\omega)$, where $A_{\mathrm{qp}}$ is the quasi-particle contribution and $A_{\mathrm{reg}}(\mathbf{k},\omega)$ is a regular background. In the limit $T\rightarrow 0$, $A_{\mathrm{qp}}$ becomes a delta-function like peak: $A_{\mathrm{qp}} \approx Z_{\mathbf{k_F}} \delta \left(\omega - \varepsilon_{\mathbf{k}} \right)$, where $\varepsilon_{\mathbf{k}} =  \mathbf{v_F} \cdot (\mathbf{k} - \mathbf{k_F})$ is the quasi-particle dispersion near the Fermi surface, $Z_{\mathbf{k_F}}$ is the quasi-particle weight, $\mathbf{v_F}$ is the Fermi velocity. At finite but small $T$ and $\varepsilon_{\mathbf{k}}$, the quasi-particle peak at $\mathbf{k_F}$ obtains a width that scales as $\max(T^2, \varepsilon_{\mathbf{k}}^2)$. The regular part satisfies $A_{\mathrm{reg}}(\mathbf{k},\omega) = O[\max(T^2,\omega^2, \varepsilon_{\mathbf{k}}^2)] $.

Inserting the Fermi liquid form of $A(\mathbf{k},\omega)$ into Eq.~(\ref{eq:G3}), we obtain that at low temperatures,
\begin{equation}
\label{eq:Zdef2}
G(\mathbf{k},\tau>0) \approx \frac{Z_{\mathbf{k_F}} e^{-\varepsilon_{\mathbf{k}}(\tau - \beta/2)}}{2 \cosh(\beta \varepsilon_{\mathbf{k}} /2)}.
\end{equation}
We can use this relation to estimate $Z_{\mathbf{k_F}}$ and $v_\mathbf{k_F}$ along the Fermi surface. For a point on the Fermi surface, $Z_{\mathbf{k_F}} = 2G(\mathbf{k_F}, \frac{\beta}{2})$.  The dispersion can be estimated from $\varepsilon_{\mathbf{k}} = -d\log[G({\mathbf{k}},\tau)] / d \tau  |_{\tau = \beta/2}$, and the Fermi velocity is then evaluated according to $v_{\mathbf{k}} =  \nabla_{\mathbf{k}} \varepsilon_\mathbf{k}$. Some caution is needed because in our finite size systems, the values of $\mathbf{k}$ are quantized, and the ${\mathbf{k}}$ grid does not in general intersect the Fermi surface. One can instead estimate $Z_{\mathbf{k}}$ from the points nearest to the Fermi surface. Also, our measurements of $G(\mathbf{k},\tau)$ are taken at finite temperatures, yielding the finite temperature estimators $\widetilde{N}(T)$, $\mathbf{v_F}(T),$  and $Z_{\bf{k_F}}(T)$  discussed in the text (Equations \ref{eq:Ntilde}, \ref{vf}, and \ref{eq:Zdef}). An extrapolation to the limit $T\rightarrow 0 $ is necessary in order to deduce the ``true" Fermi liquid properties, to the extent they are well defined. 
\section{Fermion Green's function and self-energy}
\label{app:G_and_sigma}
In Fig.~\ref{fig:matsubara_G} we show the Matsubara frequency dependence of the inverse fermion Green's function. Linear frequency dependence, consistent with a Fermi liquid, seems to dominate down to the lowest frequencies available. Finer details are visible by considering the self energy (right panels) which shows both a noticeable temperature dependence and a deviation from linearity for $\theta=0^\circ$.  
\section{Additional data}
\label{app:dump}
\begin{figure}
\includegraphics[clip=true,trim= 0 0 0 0,width= \columnwidth ]{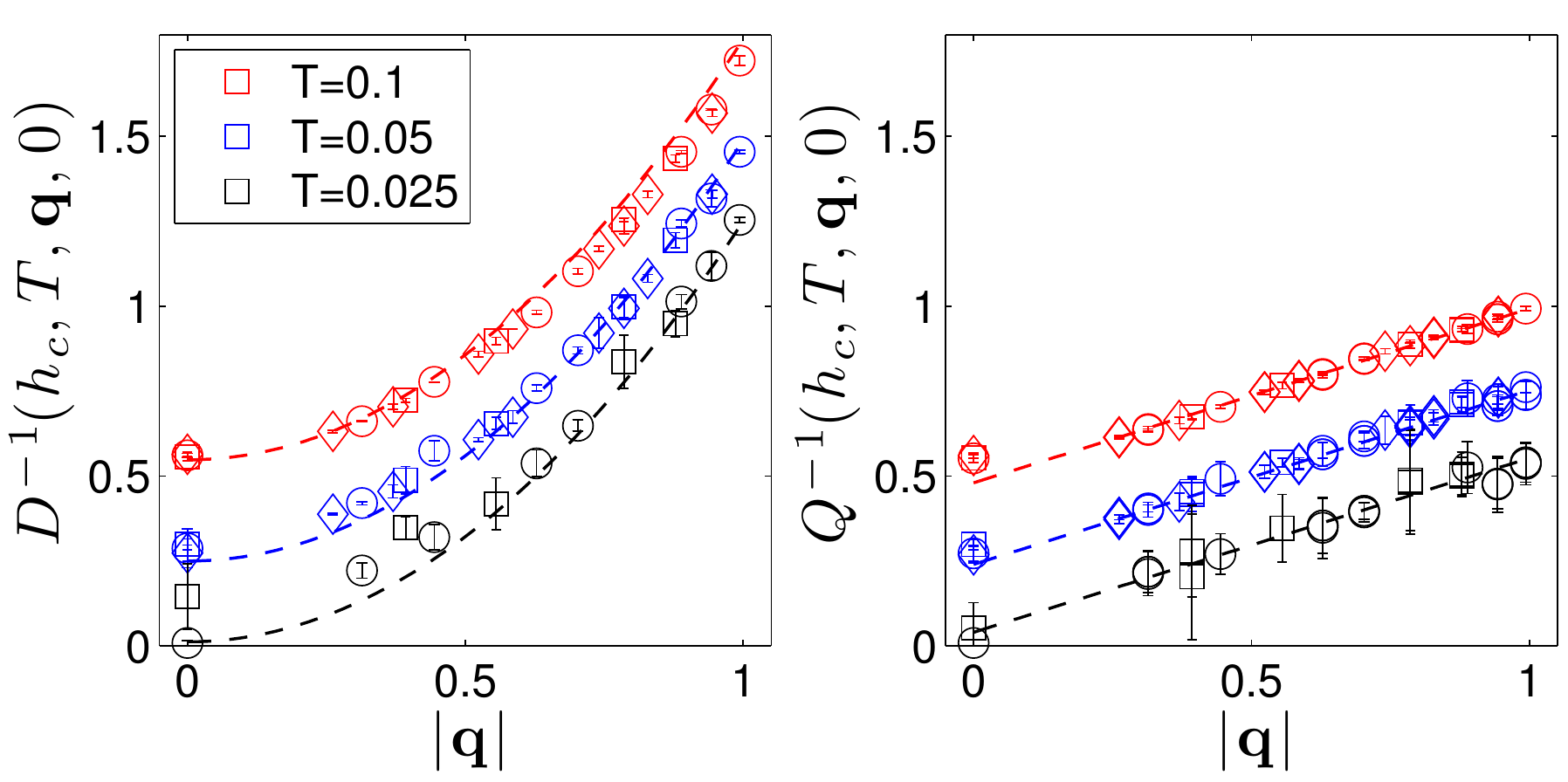}
\caption{Inverse nematic and quadrupolar correlators vs $|\mathbf{q}| $
 for all orientations of $\mathbf{q}$ at $h=h_c$ and temperatures $T=0.1,0.05,0.025$ (curves are offset for clarity).
Momenta from multiple system sizes (represented by squares for $L=16$, circles for $L=20$, and diamonds for $L=24$) fall on the same curve, indicating that the properties shown are in the thermodynamic limit. On the plot of $D^{-1}$ at left, the dotted lines are the functional approximant, Eq. \eqref{eq:D_dynamics}. On the plot of $Q^{-1}$ at right, the dotted lines are a linear fit to the data at nonzero momentum, with a temperature-independent slope. The apparent linear momentum dependence of $Q^{-1}$ suggests a critical power law, though with an apparently different exponent than the one found in $D^{-1}$.}
\label{fig:comparechiq2}
\end{figure}

Fig. \ref{fig:comparechiq} of the main text shows the momentum dependence of $D^{-1}$ and $Q^{-1}$ at $h\approx h_c$ and temperatures $T\geq 0.1 t$. The momentum dependence at low temperatures is shown in Fig. \ref{fig:comparechiq2}, here plotted versus $|\mathbf{q}|$ instead of $|\mathbf{q}|^2$. Whereas $D^{-1}$ is quadratic in $|\mathbf{q}|$, $Q^{-1}$ is apparently linear in $|\mathbf{q}|$ at low temperatures over a substantial range of momenta, a finding we have not yet understood.

\begin{figure}
\includegraphics[width=0.95\columnwidth ]{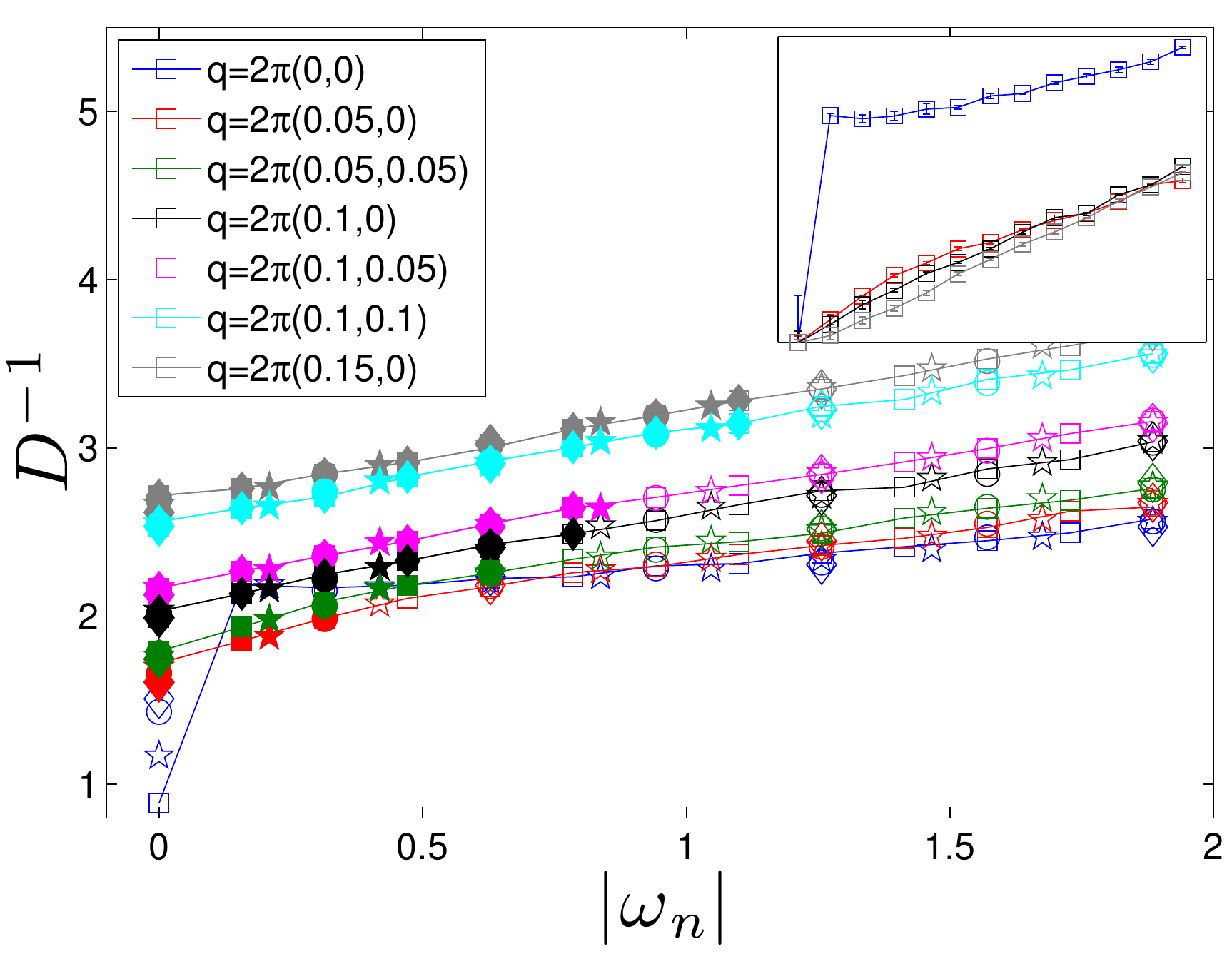}
\includegraphics[width=0.95\columnwidth ]{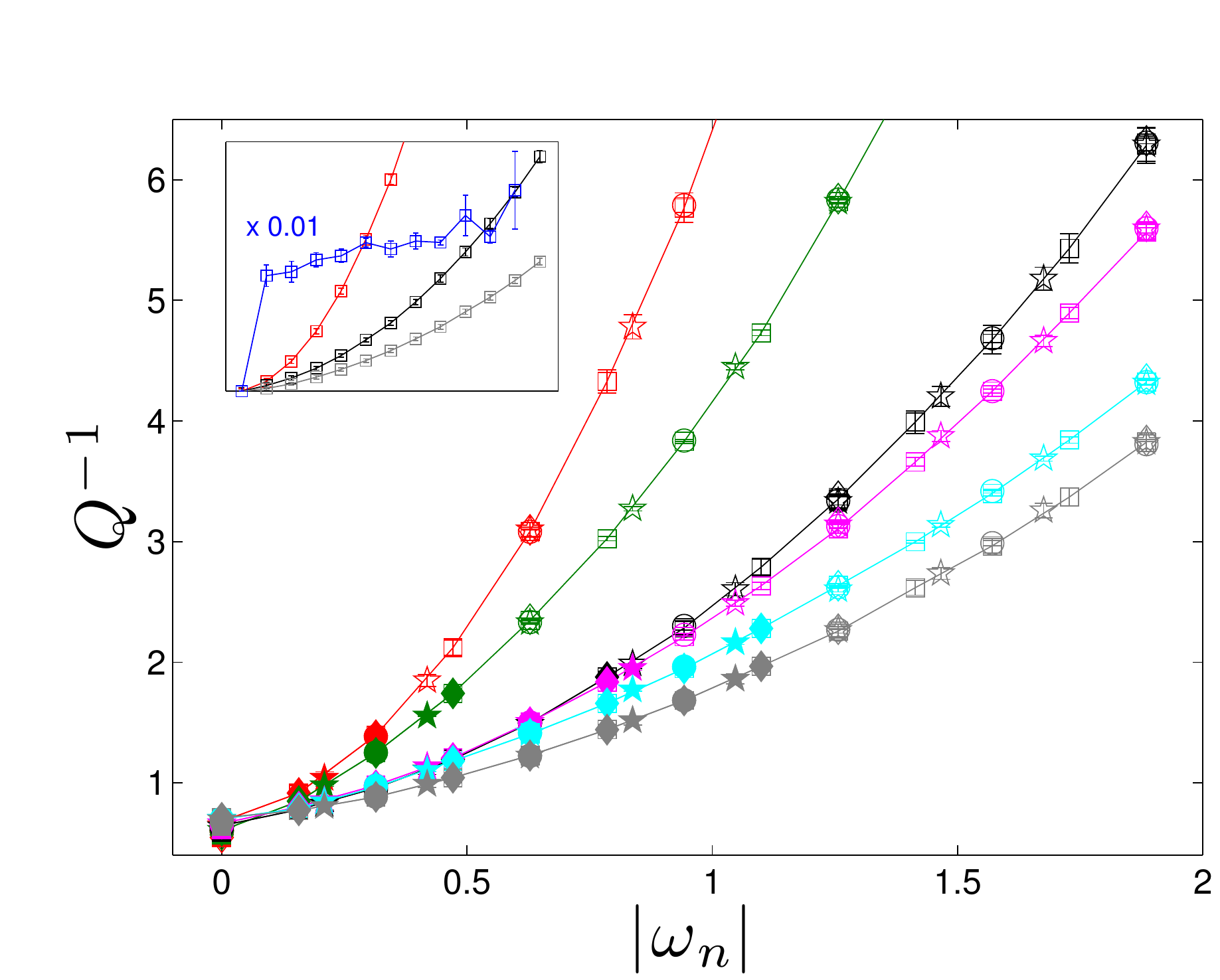}

\caption{Frequency dependence of the inverse nematic correlator $D^{-1}$ (top) and inverse quadrupolar correlator $Q^{-a}$ (bottom) at $h=4.5\approx 1.3 h_c$ for a variety of small momenta $\mathbf{q}$ in an $L=20$ system at temperatures $T=0.025, 0.033,0.05,0.1$ (shown 
as squares, stars, circles, and diamonds respectively). 
Here, there is a non-vanishing energy scale associated with nematic fluctuations since we are deep in the disordered phase, but the frequency dependence is similar to that at $h=h_c$ (see Fig. \ref{fig:chi w k}) . Different colors represent different values of $\mathbf{q}$, and filled symbols mark frequencies 
$|\omega_n| < v_F |{\bf q}|$. 
The insets show 
the subset of the data with $T=0.025$ and momenta along the $(10)$ direction
shifted by their zero frequency values. For the inset of the bottom panel the data at $\mathbf{q}=0$ are scaled down by a factor of $0.01$.  
 }
\label{fig:chiwk_gapped}
\end{figure}

Figs. \ref{fig:chi w k} and \ref{fig:chi_ff} of the main text show $D^{-1}$ and $Q^{-1}$ versus frequency for a variety of momenta, at low temperature and $h\approx h_c$. Fig. \ref{fig:chiwk_gapped} shows similar data for $h\approx 1.3 h_c$, illustrating that deviation from criticality appears mostly through an additive shift of the inverse correlator.

\begin{figure}
\includegraphics[width=0.9 \columnwidth ]{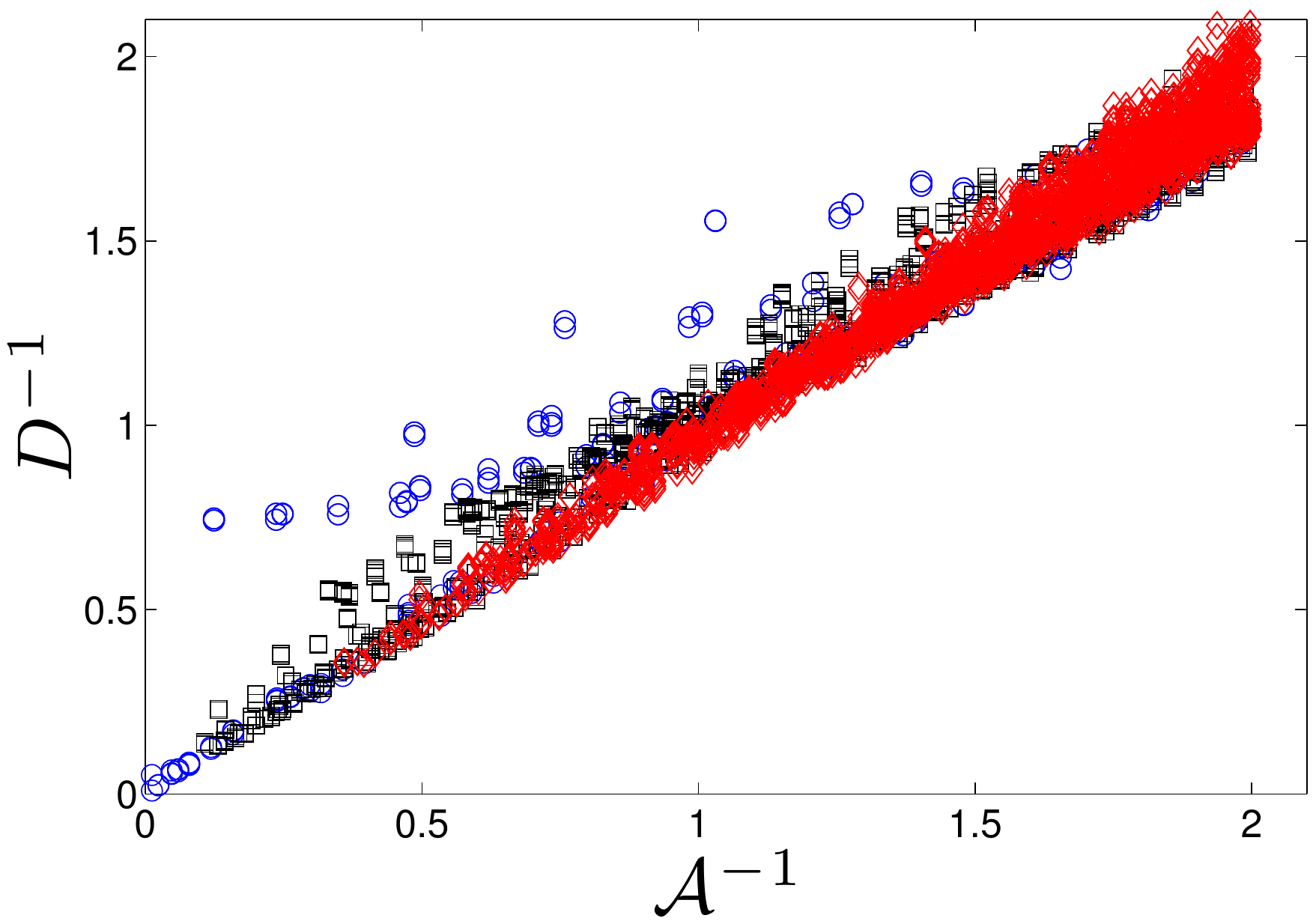}
\caption{
 Comparison between the inverse of the full dynamical functional approximant from Eq.~\ref{eq:D_dynamics}, ${\cal A}^{-1}$,  and the inverse of the nematic correlator, $D^{-1}$.
Data shown represent $L=16$, 20,  and 24, 
$h=3.525$, 3.7, 3.9, and 4.1, and 
 temperatures $T=1.0$, 0.67, 0.5, 0.33, 0.25, 0.17, 0.13, 0.1,  0.05, and 0.025. 
To exhibit the momentum dependence we have used  
blue circles for  $\mathbf{q}=\mathbf{0}$ (403 data points), black squares for $|\mathbf{q}|=(2\pi/L)$ and $(2\pi/L)\sqrt{2}$ (2,840 data points), and red diamonds for all other $\mathbf{q}$ (5,332 data points).
}
\label{fig:chi huge}
\end{figure}

Fig. \ref{fig:static} of the main text shows the validity of the functional approximant $\cal A$ of Eq. \eqref{eq:D_dynamics} in describing thermodynamic nematic correlations $D(h,T,\mathbf{q},\omega_n=0)$.  Fig. \ref{fig:chi huge} plots $D^{-1}$ vs $\cal A$ for all frequencies. The agreement is very good except for the smallest momenta, shown in blue and black.

\begin{figure}
\includegraphics[clip=true,trim= 0 0 0 0,width=\columnwidth ]{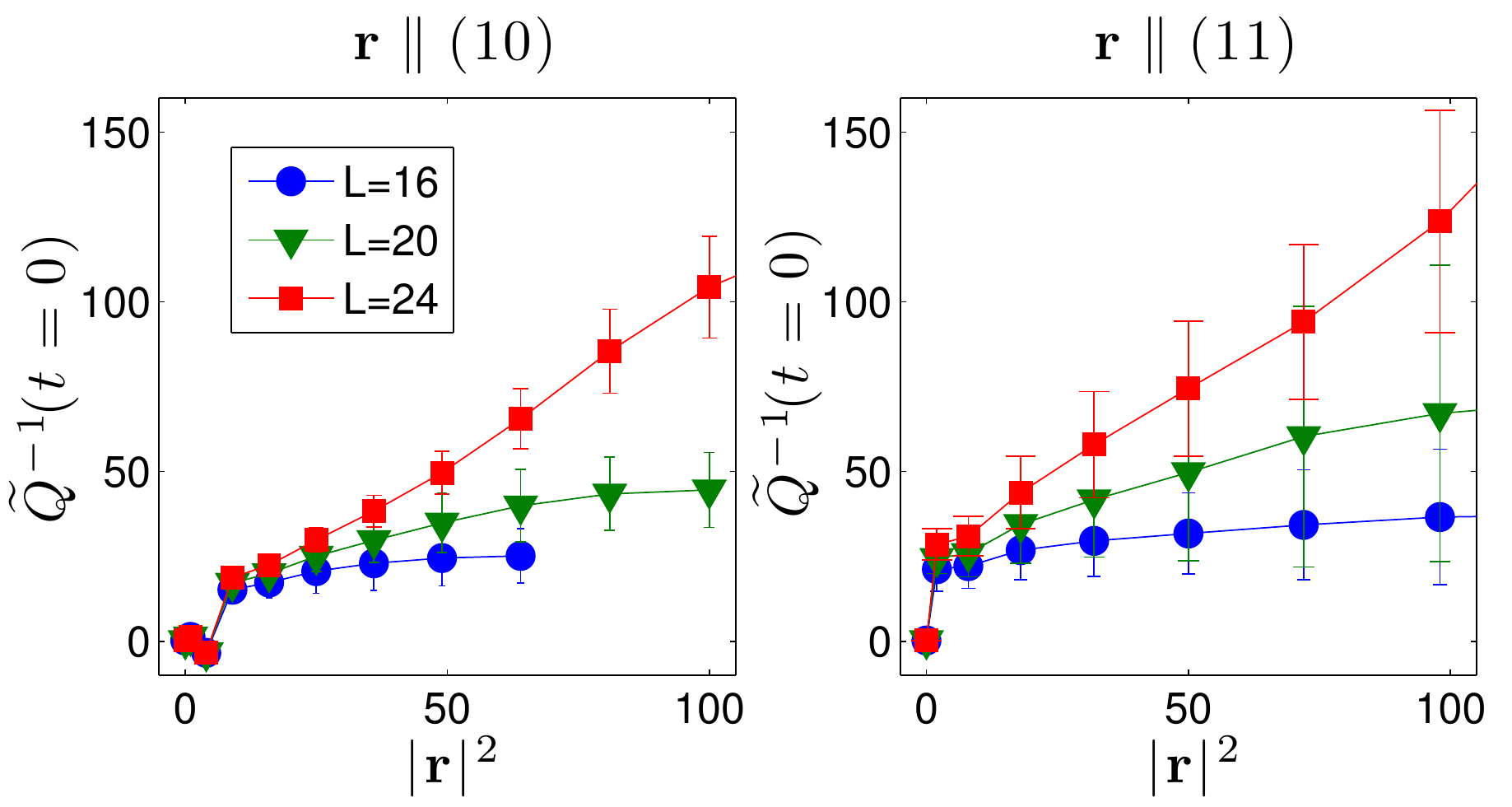}
\caption{The inverse of the equal time 
 quadrupolar correlator at $h=h_c$ and $T=0.1$, plotted versus the square of the spatial separation ${\bf r}$ along high symmetry directions for several system sizes.  In contrast to the data of Fig. \ref{fig:chi r}, substantial tetragonal anisotropy can be seen by comparing the left panel (horizontal displacements) and the right panel (diagonal displacements). However, the data apparently represent the thermodynamic limit only for $|\mathbf{r}| \lesssim 4$, and therefore we cannot reliably extract long-distance behavior.}
\label{fig:chiQ r}
\end{figure}

Fig. \ref{fig:chi r} of the main text shows the position dependence of the equal time nematic correlator $\widetilde D$. The analogous data for the equal time quadrupolar correlator $\widetilde Q$ are somewhat more messy. 
This can be seen in Fig. \ref{fig:chiQ r} where we plot the dependence of $\tilde Q on \mathbf{r}$ for $\mathbf{r}$ along the symmetry directions, $(1,0)$ and $(11)$; as is apparent, the error bars (associated with the sensitivity of the results to boundary conditions) are sufficiently large that it is difficult to conclude much from this figure.


\bibliography{}

\begin{thebibliography}{106}
\bibitem [{com()}]{comment_diamagnetic}%
  \BibitemOpen
  \href@noop {} {}\bibinfo {note} {{The paramagnetic sign of the orbital
  susceptibility is probably a band structure effect. For example, in a
  non-interacting tight-binding model on a square lattice, the orbital
  susceptibility is paramagnetic in a range of densities near half filling.
  Near the band bottom (or top), it becomes diamagnetic.}}\BibitemShut {Stop}%
\bibitem [{not()}]{note-preliminary}%
  \BibitemOpen
  \href@noop {} {}\bibinfo {note} {Preliminary results for stronger values
  $\alpha \geq 1$ of the coupling constant between the fermions and the
  critical fluctuations~\cite{Lederer-future} provide further evidence for such
  a crossover. However, at these stronger couplings we also find an intervening
  superconducting dome obscuring the QCP, which again makes the interpretation
  of these results in terms of an asymptotic scaling analysis somewhat
  problematic.}\BibitemShut {Stop}%
\end{thebibliography}

\end{document}